\documentclass[12pt]{article}
\usepackage{times}
\usepackage{geometry}
\usepackage[utf8]{inputenc}
\usepackage{enumitem,amssymb}
\usepackage[final]{pdfpages}
\usepackage{ragged2e}
\usepackage{xcolor,sectsty}
\usepackage{fancyhdr}
\usepackage{wrapfig}
\usepackage{subfigure}
\usepackage{graphbox}

\definecolor{astral}{RGB}{153,25,94}
\usepackage[colorlinks=true, citecolor=astral, linkcolor=astral, urlcolor=astral]{hyperref}

\usepackage{titlesec}
\titlespacing*{\section}{0pt}{10pt}{7pt}
\usepackage[labelfont=bf, skip=2pt]{caption}

\subsectionfont{\color{astral}}
\sectionfont{\color{astral}}
\titleformat*{\section}{\color{astral}\fontsize{15}{18}\bfseries}

\newlist{thematic}{itemize}{8}
\setlist[thematic]{label=$\square$}
\usepackage{fourier}
\usepackage{pifont}
\usepackage{lineno}
\usepackage[numbers]{natbib}
\usepackage[normalem]{ulem}
\usepackage{graphics,graphicx}
\def\apj{\rm ApJ}
\def\apjl{\rm ApJL}

\def\mnras{\rm MNRAS}
\def\araa{\rm ARAA}

\def\aap{\rm A\&A}
\def\prd{\rm PRD}
\def\prl{\rm PRL}
\def\apss{\rm Ap\&SS}
\usepackage{color}

\newcommand\partitle[1]{{\color{astral}\textbf{#1}}}

\newcommand\conclusion[1]{{\color{astral}\textbf{\large{#1}}}}

\begin{document}

\includepdf[pages=-]{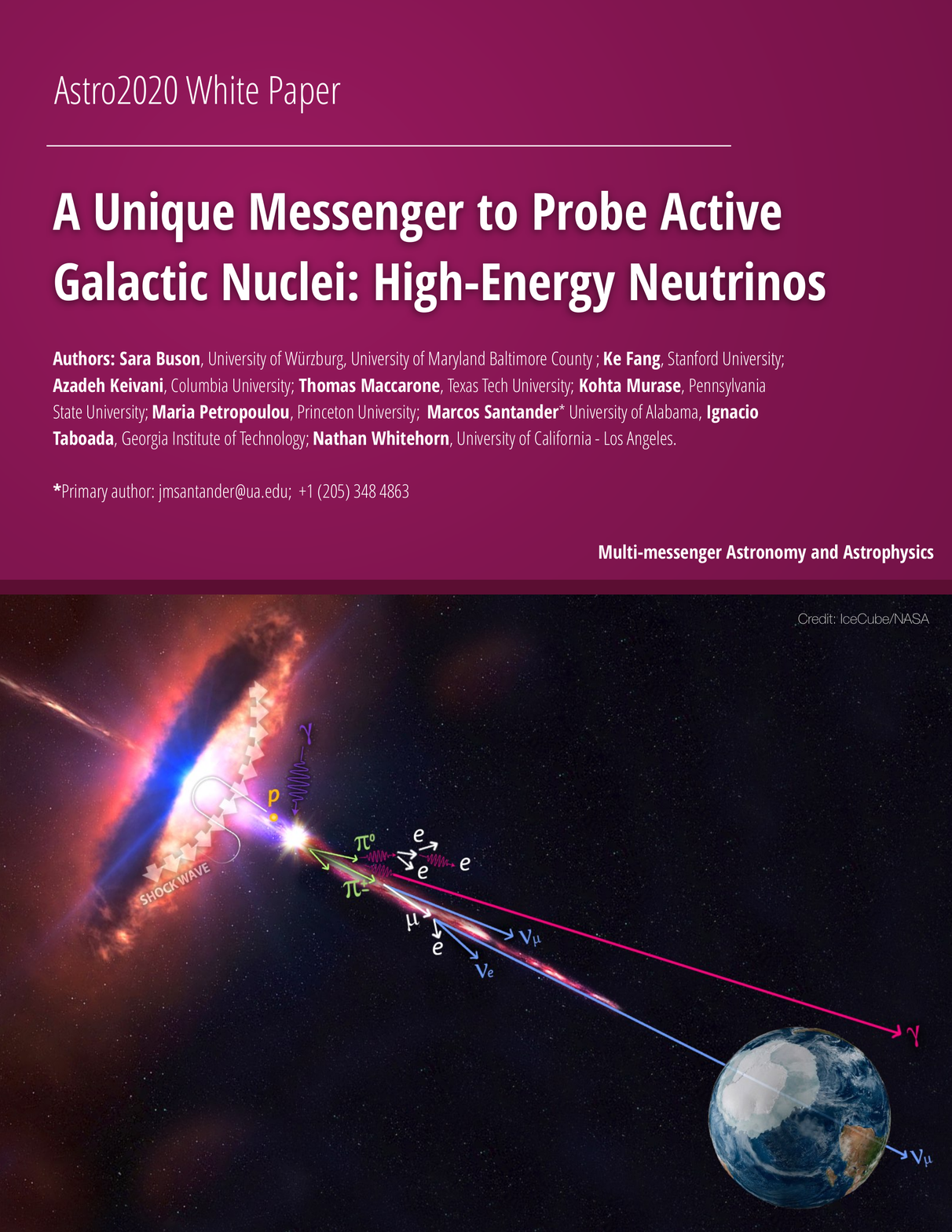} 
\clearpage
\setcounter{page}{1}

\section*{List of endorsers}
\noindent Atreya {Acharyya}\\ 
{\emph{Durham University, United Kingdom}} 
 \vspace{3pt}

\noindent Ivan {Agudo}\\ 
{\emph{IAA-CSIC, Spain}} 
 \vspace{3pt}

\noindent Juan Antonio {Aguilar S\'{a}nchez}\\ 
{\emph{Universit\'{e} Libre de Bruxelles , Belgium}} 
 \vspace{3pt}

\noindent Markus {Ahlers}\\ 
{\emph{ Niels Bohr Institute - University of Copenhagen, Denmark}} 
 \vspace{3pt}

\noindent Marco {Ajello}\\ 
{\emph{Clemson University, United States}} 
 \vspace{3pt}

\noindent Cesar {Alvarez}\\ 
{\emph{Autonomous University of Chiapas, Mexico}} 
 \vspace{3pt}

\noindent Rafael {Alves Batista}\\ 
{\emph{Universidade de Sao Paulo, Brazil}} 
 \vspace{3pt}

\noindent Karen {Andeen}\\ 
{\emph{Marquette University, United States}} 
 \vspace{3pt}

\noindent Carla {Aramo}\\ 
{\emph{INFN - Sezione di Napoli, Italy}} 
 \vspace{3pt}

\noindent Roberto {Arceo}\\ 
{\emph{Autonomous University of Chiapas, Mexico}} 
 \vspace{3pt}

\noindent Jan {Auffenberg}\\ 
{\emph{RWTH Aachen University, Germany}} 
 \vspace{3pt}

\noindent Hugo {Ayala}\\ 
{\emph{Pennsylvania State University, United States}} 
 \vspace{3pt}

\noindent Matthew {Baring}\\ 
{\emph{Rice University, United States}} 
 \vspace{3pt}

\noindent Ulisses  {Barres de Almeida}\\ 
{\emph{CBPF, Brazil}} 
 \vspace{3pt}

\noindent Imre {Bartos}\\ 
{\emph{University of Florida, United States}} 
 \vspace{3pt}

\noindent Volker {Beckmann}\\ 
{\emph{CNRS / IN2P3, France}} 
 \vspace{3pt}

\noindent Segev {BenZvi}\\ 
{\emph{University of Rochester, United States}} 
 \vspace{3pt}

\noindent Elisa {Bernardini}\\ 
{\emph{University of Padova and DESY Zeuthen, Italy and Germany}} 
 \vspace{3pt}

\noindent Ciro {Bigongiari}\\ 
{\emph{INAF - OAR, Italy}} 
 \vspace{3pt}

\noindent Erik {Blaufuss}\\ 
{\emph{ University of Maryland - College Park, United States}} 
 \vspace{3pt}

\noindent Peter {Boorman}\\ 
{\emph{University of Southampton, United Kingdom}} 
 \vspace{3pt}

\noindent Olga {Botner}\\ 
{\emph{Uppsala University, Sweden}} 
 \vspace{3pt}

\noindent Kai {Br\"{u}gge}\\ 
{\emph{TU Dortmund, Germany}} 
 \vspace{3pt}

\noindent Mauricio {Bustamante}\\ 
{\emph{ Niels Bohr Institute - University of Copenhagen, Denmark}} 
 \vspace{3pt}

\noindent Alessandro {Caccianiga}\\ 
{\emph{INAF (Istituto Nazionale di Astrofisica), Italy}} 
 \vspace{3pt}

\noindent Regina {Caputo}\\ 
{\emph{NASA GSFC, United States}} 
 \vspace{3pt}

\noindent Sylvain {Chaty}\\ 
{\emph{ University Paris Diderot - CEA Saclay, France}} 
 \vspace{3pt}

\noindent Andrew {Chen}\\ 
{\emph{University of the Witwatersrand, South Africa}} 
 \vspace{3pt}

\noindent Teddy {Cheung}\\ 
{\emph{Naval Research Lab, United States}} 
 \vspace{3pt}

\noindent Stefano {Ciprini}\\ 
{\emph{INFN Rome Tor Vergata , Italy}} 
 \vspace{3pt}

\noindent Brian {Clark}\\ 
{\emph{The Ohio State University, United States}} 
 \vspace{3pt}

\noindent Alexis {Coleiro}\\ 
{\emph{APC / Universit\'{e} Paris Diderot , France}} 
 \vspace{3pt}

\noindent Paolo {Coppi}\\ 
{\emph{Yale University, United States}} 
 \vspace{3pt}

\noindent Douglas {Cowen}\\ 
{\emph{Pennsylvania State Univeristy, United States}} 
 \vspace{3pt}

\noindent Pierre {Cristofari}\\ 
{\emph{Gran Sasso Science Institute, Italy}} 
 \vspace{3pt}

\noindent Filippo {D'Ammando}\\ 
{\emph{INAF-IRA Bologna, Italy}} 
 \vspace{3pt}

\noindent Gwenha\"{e}l {de Wasseige}\\ 
{\emph{APC - Univ Paris Diderot - CNRS/IN2P3 - CEA/Irfu - Obs de Paris - Sorbonne Paris Cit\'{e}, France}} 
 \vspace{3pt}

\noindent Cosmin {Deaconu}\\ 
{\emph{University of Chicago, United States}} 
 \vspace{3pt}

\noindent Charles {Dermer}\\ 
{\emph{Naval Research Lab, United States}} 
 \vspace{3pt}

\noindent Abhishek {Desai}\\ 
{\emph{Clemson University, United States}} 
 \vspace{3pt}

\noindent Paolo {Desiati}\\ 
{\emph{University of Wisconsin-Madison, United States}} 
 \vspace{3pt}

\noindent Tyce {DeYoung}\\ 
{\emph{Michigan State University, United States}} 
 \vspace{3pt}

\noindent Tristano {Di Girolamo}\\ 
{\emph{ University of Naples "Federico II", Italy}} 
 \vspace{3pt}

\noindent Alberto {Dominguez}\\ 
{\emph{Universidad Complutense de Madrid, Spain}} 
 \vspace{3pt}

\noindent Daniela {Dorner}\\ 
{\emph{University of W\"{u}rzburg, Germany}} 
 \vspace{3pt}

\noindent Michele {Doro}\\ 
{\emph{University and INFN Padova, Italy}} 
 \vspace{3pt}

\noindent Michael {DuVernois}\\ 
{\emph{University of Wisconsin-Madison, United States}} 
 \vspace{3pt}

\noindent Manel {Errando}\\ 
{\emph{Washington University in St Louis, United States}} 
 \vspace{3pt}

\noindent Abraham {Falcone}\\ 
{\emph{Pennsylvania State Univeristy, United States}} 
 \vspace{3pt}

\noindent Qi {Feng}\\ 
{\emph{ Barnard College - Columbia University, United States}} 
 \vspace{3pt}

\noindent Chad {Finley}\\ 
{\emph{Stockholm University, Sweden}} 
 \vspace{3pt}

\noindent Nissim {Fraija}\\ 
{\emph{National Autonomous University of Mexico, Mexico}} 
 \vspace{3pt}

\noindent Anna {Franckowiak}\\ 
{\emph{DESY Zeuthen, Germany}} 
 \vspace{3pt}

\noindent Amy {Furniss}\\ 
{\emph{California State University East Bay, United States}} 
 \vspace{3pt}

\noindent Giorgio {Galanti}\\ 
{\emph{INAF-Osservatorio Astronomico di Brera, Italy}} 
 \vspace{3pt}

\noindent Simone {Garrappa}\\ 
{\emph{DESY Zeuthen, Germany}} 
 \vspace{3pt}

\noindent Ava {Ghadimi}\\ 
{\emph{University of Alabama , United States}} 
 \vspace{3pt}

\noindent Marcello {Giroletti}\\ 
{\emph{INAF, Italy}} 
 \vspace{3pt}

\noindent Roman {Gnatyk}\\ 
{\emph{Astronomical Observatory of Taras Shevchenko National University of Kyiv, Ukraine}} 
 \vspace{3pt}

\noindent Sreetama {Goswami}\\ 
{\emph{University of Alabama, United States}} 
 \vspace{3pt}

\noindent Darren {Grant}\\ 
{\emph{Michigan State University, United States}} 
 \vspace{3pt}

\noindent Tim {Greenshaw}\\ 
{\emph{University of Liverpool, United Kingdom}} 
 \vspace{3pt}

\noindent Sylvain {Guiriec}\\ 
{\emph{GWU/NASA GSFC, United States}} 
 \vspace{3pt}

\noindent Allan {Hallgren}\\ 
{\emph{Uppsala University, Sweden}} 
 \vspace{3pt}

\noindent Lasse {Halve}\\ 
{\emph{RWTH Aachen University, Germany}} 
 \vspace{3pt}

\noindent Francis {Halzen}\\ 
{\emph{University of Wisconsin-Madison, United States}} 
 \vspace{3pt}

\noindent Elizabeth {Hays}\\ 
{\emph{NASA GSFC, United States}} 
 \vspace{3pt}

\noindent Olivier {Hervet}\\ 
{\emph{UC Santa Cruz, United States}} 
 \vspace{3pt}

\noindent Bohdan {Hnatyk}\\ 
{\emph{Astronomical Observatory of Taras Shevchenko National University of Kyiv, Ukraine}} 
 \vspace{3pt}
 
 \noindent Brian {Humensky}\\ 
{\emph{Columbia University, United States}} 
 \vspace{3pt}

\noindent Susumu {Inoue}\\ 
{\emph{RIKEN, Japan}} 
 \vspace{3pt}

\noindent Weidong {Jin}\\ 
{\emph{University of Alabama, United States}} 
 \vspace{3pt}

\noindent Matthias {Kadler}\\ 
{\emph{W\"{u}rzburg University, Germany}} 
 \vspace{3pt}

\noindent Alexander {Kappes}\\ 
{\emph{University Muenster, Germany}} 
 \vspace{3pt}

\noindent Timo {Karg}\\ 
{\emph{DESY Zeuthen, Germany}} 
 \vspace{3pt}

\noindent Albrecht {Karle}\\ 
{\emph{University of Wisconsin-Madison, United States}} 
 \vspace{3pt}

\noindent Ulrich F. {Katz}\\ 
{\emph{Friedrich-Alexander University of Erlangen-N\"{u}rnberg, Germany}} 
 \vspace{3pt}

\noindent Demos {Kazanas}\\ 
{\emph{NASA GSFC, United States}} 
 \vspace{3pt}

\noindent David {Kieda}\\ 
{\emph{University of Utah, United States}} 
 \vspace{3pt}

\noindent Spencer {Klein}\\ 
{\emph{LBNL and UC Berkeley, United States}} 
 \vspace{3pt}

\noindent Hermann {Kolanoski}\\ 
{\emph{Humboldt University Berlin , Germany}} 
 \vspace{3pt}

\noindent Marek {Kowalski}\\ 
{\emph{DESY Zeuthen, Germany}} 
 \vspace{3pt}

\noindent Michael {Kreter}\\ 
{\emph{North-West University, South Africa}} 
 \vspace{3pt}

\noindent Naoko {Kurahashi Neilson }\\ 
{\emph{Drexel University , United States}} 
 \vspace{3pt}

\noindent Jean-Philippe {Lenain}\\ 
{\emph{ Sorbonne Universit\'{e} - Universit\'{e} Paris Diderot - Sorbonne Paris Cit\'{e} - CNRS/IN2P3 - LPNHE, France}} 
 \vspace{3pt}

\noindent Hui {Li}\\ 
{\emph{LANL, United States}} 
 \vspace{3pt}

\noindent Pratik {Majumdar}\\ 
{\emph{Saha Institute of Nuclear Physics, India}} 
 \vspace{3pt}

\noindent Labani {Mallick}\\ 
{\emph{Pennsylvania State Univeristy, United States}} 
 \vspace{3pt}

\noindent Szabolcs {Marka}\\ 
{\emph{Columbia University, United States}} 
 \vspace{3pt}

\noindent Mateo {Cerruti}\\ 
{\emph{ Institut de Ci\'{e}ncies del Cosmos (ICCUB) - Universitat de Barcelona (IEEC-UB), Spain}} 
 \vspace{3pt}

\noindent Daniel {Mazin}\\ 
{\emph{ ICRR - University of Tokyo, Japan}} 
 \vspace{3pt}

\noindent Julie {McEnery}\\ 
{\emph{NASA GSFC, United States}} 
 \vspace{3pt}

\noindent Frank {McNally}\\ 
{\emph{Mercer University, United States}} 
 \vspace{3pt}

\noindent Peter {M\'{e}sz\'{a}ros}\\ 
{\emph{Pennsylvania State University, United States}} 
 \vspace{3pt}

\noindent Manuel {Meyer}\\ 
{\emph{ KIPAC - Stanford and SLAC National Accelerator Laboratory, United States}} 
 \vspace{3pt}

\noindent Teresa {Montaruli}\\ 
{\emph{University of Geneva, Switzerland}} 
 \vspace{3pt}

\noindent Reshmi {Mukherjee}\\ 
{\emph{ Barnard College - Columbia University, United States}} 
 \vspace{3pt}

\noindent Lukas {Nellen}\\ 
{\emph{ ICN - Universidad Nacional Autonoma de Mexico, Mexico}} 
 \vspace{3pt}

\noindent Anna {Nelles}\\ 
{\emph{DESY Zeuthen, Germany}} 
 \vspace{3pt}

\noindent Rodrigo {Nemmen}\\ 
{\emph{Universidade de Sao Paulo, Brazil}} 
 \vspace{3pt}

\noindent Kenny Chun Yu {Ng}\\ 
{\emph{Weizmann Institute of Science, Israel}} 
 \vspace{3pt}

\noindent Hans {Niederhausen}\\ 
{\emph{Technical University Munich, Germany}} 
 \vspace{3pt}

\noindent Daniel {Nieto}\\ 
{\emph{Universidad Complutense de Madrid, Spain}} 
 \vspace{3pt}

\noindent Kyoshi {Nishijima}\\ 
{\emph{Tokai University, Japan}} 
 \vspace{3pt}

\noindent Stephan {O'Brien}\\ 
{\emph{McGill University , Canada}} 
 \vspace{3pt}

\noindent Roopesh {Ojha}\\ 
{\emph{UMBC/NASA GSFC, United States}} 
 \vspace{3pt}

\noindent Rene {Ong}\\ 
{\emph{UCLA, United States}} 
 \vspace{3pt}

\noindent Asaf {Pe'er}\\ 
{\emph{Bar Ilan University, Israel}} 
 \vspace{3pt}

\noindent Carlos  {Perez de los Heros }\\ 
{\emph{Uppsala University , Sweden}} 
 \vspace{3pt}

\noindent Eric  {Perlman }\\ 
{\emph{Florida Institute of Technology , United States}} 
 \vspace{3pt}

\noindent Roberto {Pesce}\\ 
{\emph{Physics teacher, Italy}} 
 \vspace{3pt}

\noindent Alex {Pizzuto}\\ 
{\emph{University of Wisconsin-Madison, United States}} 
 \vspace{3pt}

\noindent Elisa {Prandini}\\ 
{\emph{University of Padova, Italy}} 
 \vspace{3pt}

\noindent John {Quinn}\\ 
{\emph{University College Dublin, Ireland}} 
 \vspace{3pt}

\noindent Bindu {Rani}\\ 
{\emph{NASA GSFC, United States}} 
 \vspace{3pt}

\noindent Ren\'{e} {Reimann}\\ 
{\emph{RWTH Aachen University, Germany}} 
 \vspace{3pt}

\noindent Elisa {Resconi}\\ 
{\emph{Technical University Munich, Germany}} 
 \vspace{3pt}

\noindent Giuseppe {Romeo}\\ 
{\emph{INAF - Osservatorio Astrofisico di Catania, Italy}} 
 \vspace{3pt}

\noindent Marco {Roncadelli}\\ 
{\emph{INFN -- Pavia, Italy}} 
 \vspace{3pt}

\noindent Iftach {Sadeh}\\ 
{\emph{DESY Zeuthen, Germany}} 
 \vspace{3pt}

\noindent Ibrahim {Safa}\\ 
{\emph{University of Wisconsin-Madison, United States}} 
 \vspace{3pt}

\noindent Narek {Sahakyan}\\ 
{\emph{ICRANet-Armenia, Armenia}} 
 \vspace{3pt}

\noindent Sourav {Sarkar}\\ 
{\emph{University of Alberta, Canada}} 
 \vspace{3pt}

\noindent Konstancja {Satalecka}\\ 
{\emph{DESY Zeuthen, Germany}} 
 \vspace{3pt}

\noindent Michael {Schimp}\\ 
{\emph{Bergische Universit\"{a}t Wuppertal, Germany}} 
 \vspace{3pt}

\noindent Fabian {Sch\"{u}ssler}\\ 
{\emph{ IRFU - CEA Paris-Saclay, France}} 
 \vspace{3pt}

\noindent David {Seckel}\\ 
{\emph{University of Delaware, United States}} 
 \vspace{3pt}

\noindent Olga {Sergijenko}\\ 
{\emph{Astronomical Observatory of Taras Shevchenko National University of Kyiv, Ukraine}} 
 \vspace{3pt}

\noindent Dennis {Soldin}\\ 
{\emph{University of Delaware, United States}} 
 \vspace{3pt}

\noindent Floyd {Stecker}\\ 
{\emph{NASA GSFC, United States}} 
 \vspace{3pt}

\noindent Thomas {Stuttard}\\ 
{\emph{Niels Bohr Institute - University of Copenhagen, Denmark}} 
 \vspace{3pt}

\noindent Fabrizio {Tavecchio}\\ 
{\emph{INAF-Osservatorio Astronomico di Brera, Italy}} 
 \vspace{3pt}

\noindent David {Thompson}\\ 
{\emph{NASA GSFC, United States}} 
 \vspace{3pt}

\noindent Kirsten {Tollefson}\\ 
{\emph{Michigan State University, United States}} 
 \vspace{3pt}

\noindent Simona {Toscano}\\ 
{\emph{Universit\'{e} Libre de Bruxelles, Belgium}} 
 \vspace{3pt}

\noindent Delia {Tosi}\\ 
{\emph{University of Wisconsin-Madison, United States}} 
 \vspace{3pt}

\noindent Gino {Tosti}\\ 
{\emph{University of Perugia, Italy}} 
 \vspace{3pt}

\noindent Sara {Turriziani}\\ 
{\emph{RIKEN, Japan}} 
 \vspace{3pt}

\noindent Nick {van Eijndhoven}\\ 
{\emph{Vrije Universiteit Brussel (IIHE-VUB), Belgium}} 
 \vspace{3pt}

\noindent Justin {Vandenbroucke}\\ 
{\emph{University of Wisconsin-Madison, United States}} 
 \vspace{3pt}

\noindent Tonia {Venters}\\ 
{\emph{NASA GSFC, United States}} 
 \vspace{3pt}

\noindent Sofia {Ventura}\\ 
{\emph{University of Siena/INFN Pisa, Italy}} 
 \vspace{3pt}

\noindent Peter {Veres}\\ 
{\emph{University of Alabama in Huntsville, United States}} 
 \vspace{3pt}

\noindent Abigail {Vieregg}\\ 
{\emph{University of Chicago, United States}} 
 \vspace{3pt}

\noindent Serguei {Vorobiov}\\ 
{\emph{University of Nova Gorica, Slovenia}} 
 \vspace{3pt}

\noindent Scott {Wakely}\\ 
{\emph{University of Chicago, United States}} 
 \vspace{3pt}

\noindent Richard {White}\\ 
{\emph{Max-Planck-Institut f\"{u}r Kernphysik, Germany}} 
 \vspace{3pt}

\noindent Christopher {Wiebusch}\\ 
{\emph{RWTH Aachen University, Germany}} 
 \vspace{3pt}

\noindent Dawn {Williams}\\ 
{\emph{University of Alabama, United States}} 
 \vspace{3pt}

\noindent Stephanie {Wissel}\\ 
{\emph{California Polytechnic State University, United States}} 
 \vspace{3pt}

\noindent Arnulfo {Zepeda}\\ 
{\emph{Cinvestav, Mexico}} 
 \vspace{3pt}

\noindent Bei {Zhou}\\ 
{\emph{The Ohio State University, United States}} 
 \vspace{3pt}

\pagebreak

\section*{Active Galactic Nuclei as Neutrino Sources}
\vspace{-0.6em}

Active galactic nuclei (AGN) with relativistic jets, powered by mass accretion onto the central supermassive black hole (SMBH) of their host galaxies, are the most powerful persistent sources of electromagnetic (EM) radiation in the Universe, with typical bolometric luminosities of $10^{43}$--$10^{48}$ erg~s$^{-1}$. The extragalactic $\gamma$-ray sky~\cite{Ackermann:2014usa} is dominated by blazars, the most extreme subclass of AGN with jets pointing close to our line of sight~\cite{Dermer:2016jmw, 2016ARA&A..54..725M}. 
Blazars can be divided into two classes: BL Lac type objects and flat spectrum radio quasars (FSRQs). 
The non-thermal radiation produced in jets spans across the EM spectrum (from radio wavelengths to TeV $\gamma$-rays) and can vary in brightness over month-long timescales or just within a few minutes \citep[e.g.,][]{cui04, aharonian07, ackermann16}. 

The broadband jet radiation generally shows two broad emission features \citep{Ulrich1997,Fossati1998}. The low-energy one, extending from radio to X-rays, is believed to originate from the synchrotron emission of relativistic electrons and positrons (henceforth, electrons) in the jet. However, the origin of the high-energy component, extending to the $\gamma$-ray band, is not well understood. \emph{Leptonic} scenarios have been put forward to explain the high-energy ``hump'' as a result of inverse Compton scattering of low-energy photons from the jet itself or from its environment (e.g., accretion disk, broad line region, or dusty torus) by relativistic electrons \citep[e.g.][]{Maraschi1992, Dermer1992, Sikora1994, Bloom1996}. 

All known processes that can accelerate electrons to relativistic energies can also act on protons and heavier ions (hadrons). In fact, the latter can reach much higher energies than electrons, because they are not as strongly affected by radiative losses \citep{2019arXiv190105164B}.
If the power carried by relativistic ions in the jet is high enough, then their radiative processes become relevant.
\emph{Lepto-hadronic} scenarios, which explain the broadband emission with both leptons and hadrons, attribute the high-energy jet emission solely to interactions involving hadrons. {These processes include proton synchrotron radiation \citep[e.g.,][]{aharonian00,muecke01,muecke03,petrodimi15, cerruti15} and intra-source \citep[e.g.,][]{mannheim93,muecke03,Dermer:2012rg, sahu13,Petropoulou:2015upa,Petropoulou2016} or intergalactic electromagnetic cascades \citep[e.g.,][]{1975Ap&SS..32..461B, Essey:2009zg,Essey:2009ju,Murase:2011cy,Oikonomou:2014xea} induced by protons via photohadronic ($p\gamma$) interactions. 
Jetted AGN are also among the most promising candidate sources of ultra-high-energy cosmic rays (UHECRs), with many 
possible acceleration sites \citep{Hillas1984,2018arXiv181206025B}, such as inner and large-scale jets with knots and shear \citep[e.g.,][]{Biermann:1987ep,Ostrowski:1998ic,Rieger:2004jz,Peer:2009vnw,Kimura:2017ubz,Caprioli:2015zka}, hot spots \citep[e.g.,][]{1990PThPh..83.1071T,Rachen:1992pg}, and radio bubbles or cocoons \cite{Berezhko:2008xn}. Neutrinos from AGN can also be produced in various sites, such as  cores~\citep[e.g.,][]{Stecker1991,AlvarezMuniz:2004uz,Stecker:2005hn, Stecker:2013fxa,Tjus:2014dna,Kimura:2014jba} and jets~\cite[e.g.,][]{Biermann:1987ep,1992A&A...260L...1M,Atoyan2001,Atoyan2004,DPM14,Murase:2014foa,Dermer:2014vaa}, or in the host galaxies~\cite{Tamborra:2014xia,Hooper:2016jls,Wang:2016vbf,Lamastra:2017iyo,Liu:2017bjr,Hooper:2018wyk}, galaxy clusters~\cite{Murase:2018iyl,Kotera:2009ms,2018NatPh..14..396F}, and intergalactic space by the interaction of escaping UHECRs from AGN with cosmic radiation fields \citep[e.g.,][]{1975Ap&SS..32..461B, Essey:2009ju,Murase:2011cy}.

Unlike photons, high-energy neutrinos can only be produced by hadronic interactions.}
\emph{The detection of AGN as neutrino point sources is therefore of paramount importance not only for understanding how the most powerful and persistent particle accelerators of the Universe work but also for unveiling the origin of UHECRs that has been a big enigma for more than fifty years.} 


\vspace{-0.6em}
\section*{The Current Multi-Messenger Picture of AGN}
\vspace{-0.6em}



\begin{figure}[t]
\centering
  \includegraphics[align=c,width=\linewidth]{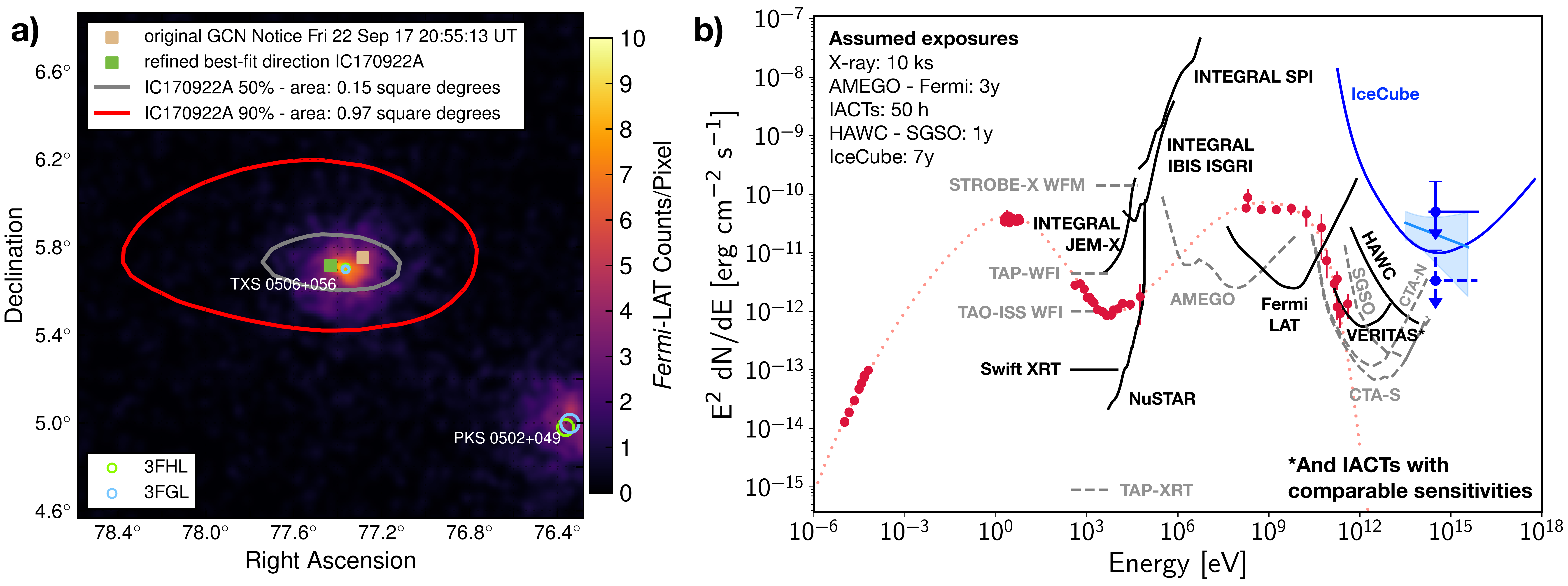}

\caption{\small{\textbf{a)} \emph{Fermi}-LAT $\gamma$-ray sky map with the error region for the IceCube-170922A event overlaid~\cite{IceCube:2018dnn}. \textbf{b)} Spectral energy distribution (SED) of TXS 0506+056 (red markers~\cite{IceCube:2018dnn}) compared to the sensitivity of current (solid black, \cite{LATSensitivity, 2015ICRC...34..771P, Abeysekara:2017mjj, DeAngelis:2017gra, NuSTAR, SwiftSensitivity}) and future (dashed gray, \cite{2017APh....93...76H, Albert:2019afb, Caputo:2017tqk, TAPSensitivity,Ray:2019pxr,2016LPICo1962.4081R}) EM instruments scaled for different exposures. Neutrino upper limits from the detection of IceCube-170922A~\cite{IceCube:2018dnn} and the best-fit neutrino spectrum from the 2014-2015 flare~\cite{IceCube:2018cha} are shown in blue compared to the seven-year sensitivity curve for IceCube~\cite{Aartsen:2016oji}. 
}}
\label{txs_map_sed}
\end{figure}

The discovery of an astrophysical neutrino flux in the 10 TeV to 10 PeV energy range by the IceCube observatory~\cite{Aartsen:2013bka,Aartsen:2013jdh} represents a breakthrough in multi-messenger astrophysics. The origin of these neutrinos remains a mystery. No strong steady~\cite{Aartsen:2016oji} or variable~\cite{Aartsen:2015wto, Aartsen:2018fpd} neutrino point sources, or a neutrino correlation with the Galactic plane~\cite{Albert:2018vxw} has been identified in the IceCube data. 
This suggests that a large population of extragalactic sources, such as non-blazar AGN, galaxy clusters/groups or star-forming galaxies, could be responsible for the bulk of the diffuse neutrino flux. 
{In addition, the similar energy densities of the diffuse neutrino and UHECR backgrounds hint at a common origin of these emissions~\cite{1999PhRvD..59b3002W}. The diffuse gamma-ray and neutrino backgrounds can also be explained simultaneously~\cite{Murase:2013rfa},  \citep{Murase:2016gly,2018NatPh..14..396F}, 
which may be explained by AGN embedded in galaxy clusters/groups or starburst galaxies~\cite{Murase:2016gly,2018NatPh..14..396F,Liu:2017bjr}. 
Nonetheless, the diffuse flux between $10-100$~TeV cannot be solely explained by either $pp$ scenarios for star-forming galaxies or $p\gamma$ scenarios for AGN jets (including blazars and radio galaxies  \citep[e.g.,][]{Murase:2014foa,Murase:2015xka,Bechtol:2015uqb,Xiao:2016rvd,Padovani:2015mba,Palladino:2018bqf}; see \cite{Murase2017} for a review). 
The contribution of $\gamma$-ray blazars, in particular, to the diffuse neutrino flux has been constrained to the level of $\sim10-30\%$ by correlation and stacking analyses
\citep{Aartsen:2016lir,2017arXiv171001179I,Murase:2018iyl}. 
The dominant contribution to the diffuse neutrino flux in the 10-100 TeV range may come from sources that are either genuinely opaque to $\gamma$-rays, such as AGN cores \citep{Kimura:2014jba} or that are hidden to current $\gamma$-ray detectors, such as MeV blazars \citep{Murase:2015xka}.
The fact that $\gamma$-ray-emitting AGN are not the dominant contributors to the bulk of the diffuse neutrino flux does not prevent them from being detectable point neutrino sources.
Several studies claimed a connection between individual $\gamma$-ray blazars and high-energy neutrino events, although with marginal correlation significances~\cite{Padovani2014, Padovani2016,Kadler:2016ygj}.  The first compelling evidence for the identification of an astrophysical high-energy neutrino source was provided in 2017 by the detection of a high-energy neutrino event (IceCube-170922A) in coincidence with a strong EM flare of the $\gamma$-ray blazar TXS 0506+056 (Fig.\ref{txs_map_sed}a)~\cite{IceCube:2018dnn}. In fact, blazar $\gamma$-ray flares are ideal periods for the detection of high-energy neutrinos due to the lower atmospheric neutrino background contamination and the higher neutrino production efficiency \citep[e.g.][]{Dermer:2014vaa,Petropoulou2016,Kadler:2016ygj, Guepin2017,Murase:2018iyl}. 
The detection of IceCube-170922A and the prompt dissemination of the neutrino sky position to the astronomical community triggered an extensive multi-messenger campaign to characterize the source emission \cite{Albert:2018kjg,Ahnen:2018mvi,Abeysekara:2018oub,Keivani:2018rnh}. 
The rich multi-wavelength data set enabled for the first time detailed theoretical modeling that could explain the neutrino emission in coincidence with the EM blazar flare~\cite{Ahnen:2018mvi,Keivani:2018rnh,Murase:2018iyl,Cerruti:2018tmc,Gao:2018mnu}.


A follow-up analysis of archival IceCube neutrino data also unveiled neutrino activity during a $\sim$100-day window in 2014-15~\cite{IceCube:2018cha}. 
Intriguingly, this detection was not accompanied by flaring in $\gamma$-rays as in the case of IceCube-170922A, although some debate exists about a potential hardening in the blazar $\gamma$-ray spectrum during the neutrino activity period~\cite{Padovani:2018acg, Aartsen:2019gxs}.
The lack of sensitive multi-wavelength observations during this period is a significant hurdle in the multi-messenger modeling of the neutrino ``flare''~\cite{Murase:2018iyl,Rodrigues:2018tku,Reimer:2018vvw}. This is particularly true for the keV to MeV band where no observations are available, but a high photon flux due to the cascade of the hadronically-produced $\gamma$-rays is theoretically expected \citep{Petropoulou:2015upa,PetroMast2015, Murase:2018iyl}.

So far, there is no convincing theoretical explanation for all multi-messenger observations of TXS~0506+056, which has raised a number of important questions:
What makes its 2014-15 flare activity special?
Is there more than one neutrino production sites in AGN?
Can we find more robust AGN-neutrino associations? 
What would be the best observing strategy,
especially if GeV $\gamma$-rays and TeV-PeV neutrinos are not produced at the same time?
We outline next the required observational capabilities to address these questions in the coming decade.


\vspace{-0.6em}
\section*{Multi-messenger Studies of AGN in the Next Decade}
\vspace{-0.6em}

The construction of next-generation neutrino telescopes coupled with an expansion of multi-wavelength follow-up efforts and the improvements in broad-band coverage and sensitivity of new EM observatories will provide a major boost in the identification and study of AGN as neutrino emitters. 
We here outline a number of activities that will help solidify the AGN high-energy neutrino connection by detecting more sources beyond TXS 0506+056. Together with multi-wavelength follow-up campaigns~\cite{Santander:2016bvv}, we will be able to probe the physics of neutrino and EM emission in AGN.  


\begin{wrapfigure}{r}{0.45\linewidth}
\vspace{-20pt}
\centering
\includegraphics[width=0.45\textwidth]{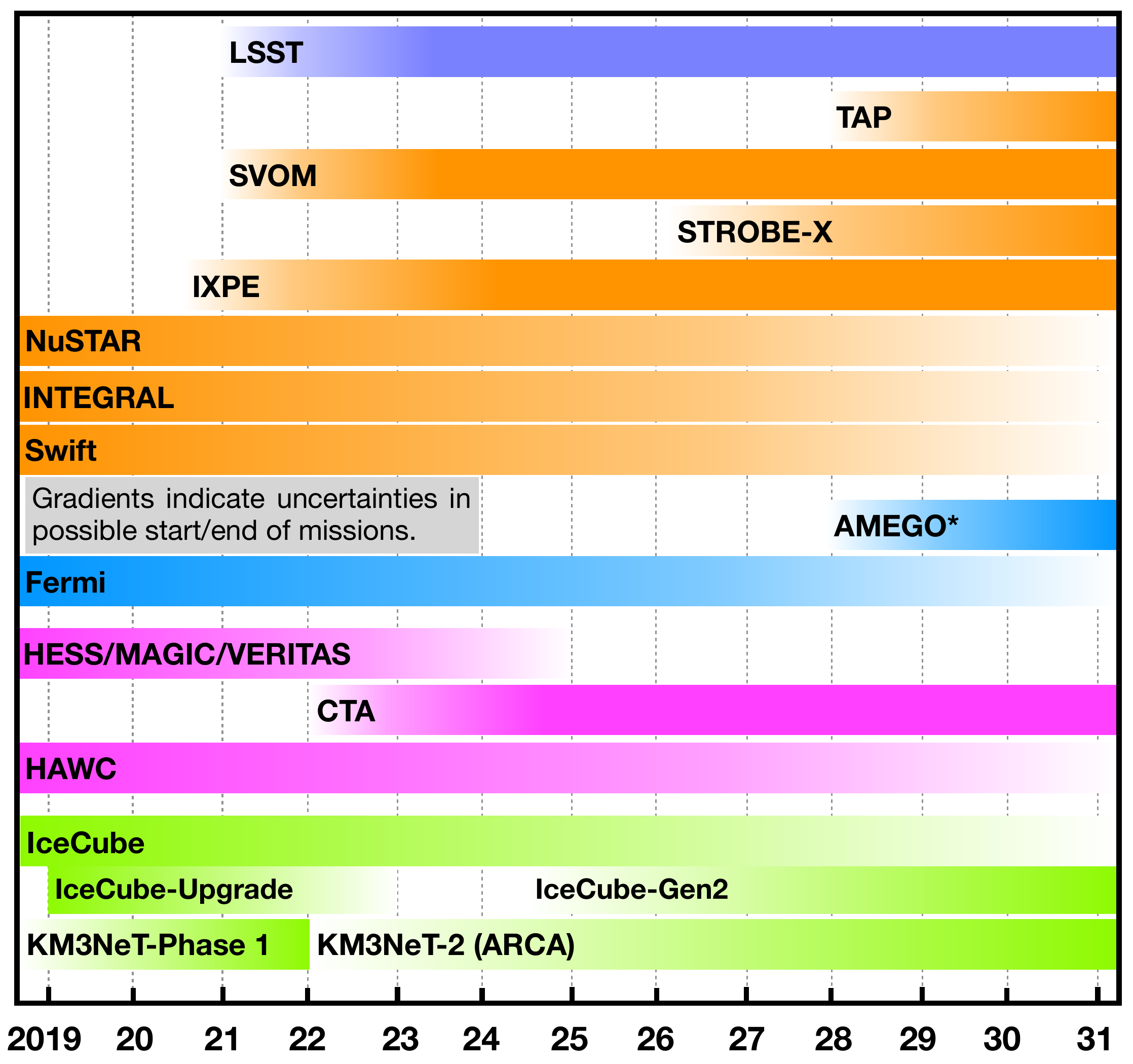}
\caption{\small{Timeline of some of the instruments expected to be involved in multi-messenger studies of AGN in the coming decade (some not yet funded or with unclear timelines).}}
\vspace{-0.15em}
\label{landscape}
\end{wrapfigure}

\noindent{\partitle{Neutrino observatories}:} 
The primary backgrounds to the detection 
of astrophysical neutrinos
are muons and neutrinos produced by cosmic ray interactions in the upper
atmosphere. These have a steeply-falling energy spectrum, with atmospheric neutrinos becoming sub-dominant to the observed astrophysical ones at $\sim100$~TeV. As a result, the primary target energy range for detection of neutrinos from AGN is in the 100~TeV--PeV range, although clustering in time or space can significantly lower the energy threshold.
The highest possible neutrino flux from UHECR sources has been calculated assuming a calorimetric relationship \cite{1999PhRvD..59b3002W}, which establishes that a gigaton or larger scale instrument is needed to observe astrophysical neutrinos above 100 TeV.

IceCube is the largest operating neutrino instrument in this energy range and the first to reach a gigaton mass. It uses the under-water/ice Cherenkov technique in the south polar ice cap achieving an angular resolution of $\lesssim 0.5^{\rm o}$, and continuously observes the entire sky. IceCube's realtime alert program notifies the astronomical community if a likely astrophysical neutrino signal is identified to enable follow-up EM observations. This includes near-realtime public alerts for single neutrinos events of likely astrophysical origin such as IceCube-170922A using the GRB Coordinate Network (GCN)~\cite{Aartsen:2016lmt}. Two underwater neutrino detectors are currently in operation in the northern hemisphere, with better angular resolution than IceCube, but much smaller volumes and thus reduced sensitivity: ANTARES~\cite{2011NIMPA.656...11A} and Baikal NT-200~\cite{Aynutdinov:2009zzb}.

The next decade will see the design, construction and operation of next-generation under-water/ice neutrino telescopes which will expand upon currently running experiments: KM3NeT \cite{Adrian-Martinez:2016fdl} and GVD~\cite{Avrorin:2018ijk} in the northern hemisphere, and IceCube-Gen2 at the South Pole~\cite{Aartsen:2014njl}. The ARCA component of KM3NeT~\cite{2018arXiv181008499T} will have a sensitivity similar to or better than that of IceCube by a factor of two. And, as a result of its mid-latitude location, this sensitivity will cover a wider range of declinations. In IceCube, best sensitivity is achieved for $\delta = -5^\circ$ to $90^\circ$, while KM3NeT will cover $\sim$95\% of the entire sky. 
The IceCube-Gen2 upgrade will increase the size of the detector by a factor of $\sim6$ and improve on sensitivity to point sources, such as AGNs, by a factor of $\sim$5 with respect to IceCube. Assuming an Euclidean geometry and uniform source distribution (admittedly simplistic), this improvement would result in $\sim$10 observations similar to that of TXS~0506+056 over 10 years with Gen2. Given their increase in sensitivity, future neutrino detectors are also expected to provide a rate of neutrino alerts substantially larger than the current $\sim 10$ per year, with minute latency, improved angular resolution ($\sim0.2^{\circ}$~\cite{vanSanten:2017chb,2018arXiv181008499T}) and higher astrophysical purity to enhance EM counterpart searches.

At $>$ 10 PeV energies, radio neutrino detectors such as the proposed ARA \citep{2012APh....35..457A} and ARIANNA arrays \citep{2017APh....90...50B} (which have recently joined efforts to propose the Radio Neutrino Observatory, RNO, in Antarctica), and GRAND \citep{2018arXiv181009994G} will characterize the high-energy end of the astrophysical neutrino spectrum and potentially identify AGN counterparts to neutrino events\footnote{A separate white paper~\cite{nuwp} details plans for high-energy astrophysical neutrino studies in the coming decade.}.

\vspace{0.2em}
\noindent{\partitle{EM Observatories}:}
Decoding the information simultaneously carried by the neutrino and EM signals is crucial for unequivocally pinpointing the production sites of multi-messenger emissions in AGN. This is not a simple task, as uniquely illustrated by the multi-messenger observations of TXS~0506+056, especially because the properties of the physical engine can vary on timescales from minutes to months.
With the advent of neutrino detectors and future EM observatories with wider field and energy-range coverage (see Figs.~\ref{txs_map_sed}b and \ref{landscape} for coverage and timeline), we will be able to test if neutrinos are correlated with periods of flaring activity in a specific energy band. Identifying such a correlation (or the lack of one) would shed light on the properties and location of the emission region.

%

EM observations of the low-energy SED ``hump'' (in the radio to X-ray range) can constrain the synchrotron emission from the AGN, which is expected to be dominated by leptonic processes. Radio observations can provide photometric coverage of a selection of radio-loud AGN \cite{Richards:2010su} and also imaging of the jet or the core regions~\cite{Lister:2008bw} that could then be correlated to a neutrino emission period~\cite{Kadler:2016ygj, Kun:2018zin}. Future facilities such as ngVLA~\cite{2018SPIE10700E..1OS} would improve on these efforts. When not affected by light constraints, optical facilities can provide sensitive monitoring of AGN across the entire sky. With several survey instruments coming online in the next decade that can provide AGN monitoring with high sensitivity and cadence, in particular LSST~\cite{Robertson:2017glv, 2009arXiv0912.0201L}, there will be many opportunities for neutrino correlation studies. 

The critical energy band for the multi-messenger modeling of AGN emission is at high energies (keV and above), where photons from hadronic processes are expected to be produced via synchrotron radiation of protons or/and secondary pairs produced by pion and muon decays or by the $\gamma\gamma$ absorption of high-energy photons~\cite{Boettcher:2013wxa, PetroMast2015, Petropoulou:2015upa, PetroVas17, Gao:2018mnu, Keivani:2018rnh}. 
In the soft X-ray band ($<10$ keV) the Neil Gehrels \emph{Swift} observatory has been the main follow-up instrument to search for EM counterparts to singlets~\cite{Evans:2015qia, Keivani:2017icrc} or multiplets~\cite{Aartsen:2017snx} of high-energy neutrinos given its rapid repointing capability. 
In addition, a \emph{Swift} monitoring program exists for \emph{Fermi}-detected sources which provides coverage of some of the brightest AGN, but a higher cadence on a larger number of sources would be desirable in the coming decade. Current wide-field instruments such as INTEGRAL~\cite{2014NIMPA.742...47U} and MAXI/GSC~\cite{Kawamuro:2018eky} offer larger sky coverage than \emph{Swift} at the expense of sensitivity, but future wide-field instruments such as TAP~\cite{2018AAS...23112105C}, STROBE-X WFM~\cite{Ray:2018dlb} (both recently selected for NASA probe mission studies) and TAO-ISS~\cite{2016LPICo1962.4081R}  would deliver competitive sensitivity while covering a large fraction of the sky. 
This capability may soon be crucial as \emph{Swift} could cease operations and other instruments like \emph{Chandra} are not well-suited for prompt observations of large sky regions. We therefore advocate the continuation of \emph{Swift} to provide soft X-ray coverage for these studies until comparable capabilities become available or are complemented by European-led missions such as SVOM~\cite{2015arXiv151203323C}.

In the hard X-ray band ($>10$ keV), \emph{NuSTAR} will continue to be the most sensitive instrument. While the observational constraints and field of view of \emph{NuSTAR} would not allow it to search for potential AGN neutrino counterparts, it could be used for follow-up observations like in the case of TXS 0506+056~\cite{2017ATel10845....1F}. No facilities sensitive enough to detect a substantial number of AGN in the MeV band currently exist, which is critical towards understanding the hadronic emission from AGN jets as the cascading of high-energy photons would results in a high flux in the hard X-ray to MeV band. Missions like AMEGO~\cite{Moiseev:2017mxg} or the European-led e-ASTROGAM~\cite{DeAngelis:2017gra} would be critical in enabling these studies. Beyond the MeV AGN monitoring, AMEGO will also provide polarimetric measurements which can help differentiate between leptonic and hadronic emission processes~\cite{Zhang:2013bna, Paliya:2018wgu}. Similar polarization signatures in the optical~\cite{Zhang:2018xrr} could be explored using existing capabilities, or with IXPE~\cite{WEISSKOPF20161179} in the X-ray range~\cite{Zhang:2013bna}.
In the GeV band, the \emph{Fermi}-LAT~\cite{2009ApJ...697.1071A} is a critical instrument to study the $\gamma$-ray emission from AGN and no comparable missions are foreseen in the coming future in this band. We therefore advocate the continuation of the \emph{Fermi} mission into the coming decade.

Current and new observatories in the very-high-energy band (VHE, $E>100$ GeV) will continue follow-up observations of neutrino events and potential AGN neutrino counterparts in the coming decade. 
Current telescopes such as H.E.S.S., MAGIC, and VERITAS will continue their neutrino follow-up programs~\cite{Santander:2017zkl} during the first half of the decade at which point it is expected that CTA will start scientific operations and provide the most sensitive coverage in the VHE band~\cite{Acharya:2017ttl}. Wide-field VHE instruments such as HAWC~\cite{Abeysekara:2017hyn}, while less sensitive than CTA, will continue to monitor a large number of AGN that could be correlated with neutrino observations. Future observatories of this type are under construction~\cite{Chen:2018glo}, and some have been proposed in the southern hemisphere where no instrumentation of this kind currently exists~\cite{Albert:2019afb, Assis:2017zzn, Thoudam:2017hgj, Ohnishi:2017qsz}. We encourage VHE $\gamma$-ray studies of AGN-neutrino correlations.

\vspace{5pt}

\noindent \conclusion{Conclusion and outlook}: The detection of astrophysical neutrinos by IceCube and the evidence for neutrino emission from a blazar offer exciting opportunities for the study of high-energy neutrinos and photons from AGN in the coming decade. We advocate for a multi-messenger approach that combines high-energy neutrino observations performed by telescopes that will come online in the next decade, and  multi-wavelength EM observations by existing and future instruments, with an emphasis on soft X-ray to VHE $\gamma$-ray coverage. The unique capabilities of these instruments combined, promise to solve several long-standing issues in our understanding of AGN, the most powerful and  persistent cosmic accelerators.

\pagebreak


\begin{thebibliography}{149}%
\makeatletter
\providecommand \@ifxundefined [1]{%
 \@ifx{#1\undefined}
}%
\providecommand \@ifnum [1]{%
 \ifnum #1\expandafter \@firstoftwo
 \else \expandafter \@secondoftwo
 \fi
}%
\providecommand \@ifx [1]{%
 \ifx #1\expandafter \@firstoftwo
 \else \expandafter \@secondoftwo
 \fi
}%
\providecommand \natexlab [1]{#1}%
\providecommand \enquote  [1]{``#1''}%
\providecommand \bibnamefont  [1]{#1}%
\providecommand \bibfnamefont [1]{#1}%
\providecommand \citenamefont [1]{#1}%
\providecommand \href@noop [0]{\@secondoftwo}%
\providecommand \href [0]{\begingroup \@sanitize@url \@href}%
\providecommand \@href[1]{\@@startlink{#1}\@@href}%
\providecommand \@@href[1]{\endgroup#1\@@endlink}%
\providecommand \@sanitize@url [0]{\catcode `\\12\catcode `\$12\catcode
  `\&12\catcode `\#12\catcode `\^12\catcode `\_12\catcode `\%12\relax}%
\providecommand \@@startlink[1]{}%
\providecommand \@@endlink[0]{}%
\providecommand \url  [0]{\begingroup\@sanitize@url \@url }%
\providecommand \@url [1]{\endgroup\@href {#1}{\urlprefix }}%
\providecommand \urlprefix  [0]{URL }%
\providecommand \Eprint [0]{\href }%
\providecommand \doibase [0]{https://doi.org/}%
\providecommand \selectlanguage [0]{\@gobble}%
\providecommand \bibinfo  [0]{\@secondoftwo}%
\providecommand \bibfield  [0]{\@secondoftwo}%
\providecommand \translation [1]{[#1]}%
\providecommand \BibitemOpen [0]{}%
\providecommand \bibitemStop [0]{}%
\providecommand \bibitemNoStop [0]{.\EOS\space}%
\providecommand \EOS [0]{\spacefactor3000\relax}%
\providecommand \BibitemShut  [1]{\csname bibitem#1\endcsname}%
\let\auto@bib@innerbib\@empty
\bibitem [{\citenamefont {Ackermann}\ \emph {et~al.}(2015)\citenamefont
  {Ackermann} \emph {et~al.}}]{Ackermann:2014usa}%
  \BibitemOpen
  \bibfield  {author} {\bibinfo {author} {\bibfnamefont {M.}~\bibnamefont
  {Ackermann}} \emph {et~al.} (\bibinfo {collaboration} {Fermi-LAT}),\ }\href
  {https://doi.org/10.1088/0004-637X/799/1/86} {\bibfield  {journal} {\bibinfo
  {journal} {Astrophys. J.}\ }\textbf {\bibinfo {volume} {799}},\ \bibinfo
  {pages} {86} (\bibinfo {year} {2015})},\ \Eprint
  {https://arxiv.org/abs/1410.3696} {arXiv:1410.3696 [astro-ph.HE]}
  \BibitemShut {NoStop}%
\bibitem [{\citenamefont {Dermer}\ and\ \citenamefont
  {Giebels}(2016)}]{Dermer:2016jmw}%
  \BibitemOpen
  \bibfield  {author} {\bibinfo {author} {\bibfnamefont {C.~D.}\ \bibnamefont
  {Dermer}}\ and\ \bibinfo {author} {\bibfnamefont {B.}~\bibnamefont
  {Giebels}},\ }\href {https://doi.org/10.1016/j.crhy.2016.04.004} {\bibfield
  {journal} {\bibinfo  {journal} {Comptes Rendus Physique}\ }\textbf {\bibinfo
  {volume} {17}},\ \bibinfo {pages} {594} (\bibinfo {year} {2016})},\ \Eprint
  {https://arxiv.org/abs/1602.06592} {arXiv:1602.06592 [astro-ph.HE]}
  \BibitemShut {NoStop}%
\bibitem [{\citenamefont {{Madejski}}\ and\ \citenamefont
  {{Sikora}}(2016)}]{2016ARA&A..54..725M}%
  \BibitemOpen
  \bibfield  {author} {\bibinfo {author} {\bibfnamefont {G.}~\bibnamefont
  {{Madejski}}}\ and\ \bibinfo {author} {\bibfnamefont {M.}~\bibnamefont
  {{Sikora}}},\ }\href {https://doi.org/10.1146/annurev-astro-081913-040044}
  {\bibfield  {journal} {\bibinfo  {journal} {\araa}\ }\textbf {\bibinfo
  {volume} {54}},\ \bibinfo {pages} {725} (\bibinfo {year} {2016})}\BibitemShut
  {NoStop}%
\bibitem [{\citenamefont {{Cui}}(2004)}]{cui04}%
  \BibitemOpen
  \bibfield  {author} {\bibinfo {author} {\bibfnamefont {W.}~\bibnamefont
  {{Cui}}},\ }\href {https://doi.org/10.1086/382587} {\bibfield  {journal}
  {\bibinfo  {journal} {\apj}\ }\textbf {\bibinfo {volume} {605}},\ \bibinfo
  {pages} {662} (\bibinfo {year} {2004})},\ \Eprint
  {https://arxiv.org/abs/astro-ph/0401222} {arXiv:astro-ph/0401222 [astro-ph]}
  \BibitemShut {NoStop}%
\bibitem [{\citenamefont {{Aharonian}}\ \emph {et~al.}(2007)\citenamefont
  {{Aharonian}} \emph {et~al.}}]{aharonian07}%
  \BibitemOpen
  \bibfield  {author} {\bibinfo {author} {\bibfnamefont {F.}~\bibnamefont
  {{Aharonian}}} \emph {et~al.},\ }\href {https://doi.org/10.1086/520635}
  {\bibfield  {journal} {\bibinfo  {journal} {\apjl}\ }\textbf {\bibinfo
  {volume} {664}},\ \bibinfo {pages} {L71} (\bibinfo {year} {2007})},\ \Eprint
  {https://arxiv.org/abs/0706.0797} {arXiv:0706.0797} \BibitemShut {NoStop}%
\bibitem [{\citenamefont {{Ackermann}}\ \emph {et~al.}(2016)\citenamefont
  {{Ackermann}} \emph {et~al.}}]{ackermann16}%
  \BibitemOpen
  \bibfield  {author} {\bibinfo {author} {\bibfnamefont {M.}~\bibnamefont
  {{Ackermann}}} \emph {et~al.} (\bibinfo {collaboration} {Fermi-LAT}),\ }\href
  {https://doi.org/10.3847/2041-8205/824/2/L20} {\bibfield  {journal} {\bibinfo
   {journal} {\apjl}\ }\textbf {\bibinfo {volume} {824}},\ \bibinfo {eid} {L20}
  (\bibinfo {year} {2016})},\ \Eprint {https://arxiv.org/abs/1605.05324}
  {arXiv:1605.05324 [astro-ph.HE]} \BibitemShut {NoStop}%
\bibitem [{\citenamefont {{Ulrich}}\ \emph {et~al.}(1997)\citenamefont
  {{Ulrich}}, \citenamefont {{Maraschi}},\ and\ \citenamefont
  {{Urry}}}]{Ulrich1997}%
  \BibitemOpen
  \bibfield  {author} {\bibinfo {author} {\bibfnamefont {M.-H.}\ \bibnamefont
  {{Ulrich}}}, \bibinfo {author} {\bibfnamefont {L.}~\bibnamefont
  {{Maraschi}}},\ and\ \bibinfo {author} {\bibfnamefont {C.~M.}\ \bibnamefont
  {{Urry}}},\ }\href {https://doi.org/10.1146/annurev.astro.35.1.445}
  {\bibfield  {journal} {\bibinfo  {journal} {\araa}\ }\textbf {\bibinfo
  {volume} {35}},\ \bibinfo {pages} {445} (\bibinfo {year} {1997})}\BibitemShut
  {NoStop}%
\bibitem [{\citenamefont {{Fossati}}\ \emph {et~al.}(1998)\citenamefont
  {{Fossati}}, \citenamefont {{Maraschi}}, \citenamefont {{Celotti}},
  \citenamefont {{Comastri}},\ and\ \citenamefont
  {{Ghisellini}}}]{Fossati1998}%
  \BibitemOpen
  \bibfield  {author} {\bibinfo {author} {\bibfnamefont {G.}~\bibnamefont
  {{Fossati}}}, \bibinfo {author} {\bibfnamefont {L.}~\bibnamefont
  {{Maraschi}}}, \bibinfo {author} {\bibfnamefont {A.}~\bibnamefont
  {{Celotti}}}, \bibinfo {author} {\bibfnamefont {A.}~\bibnamefont
  {{Comastri}}},\ and\ \bibinfo {author} {\bibfnamefont {G.}~\bibnamefont
  {{Ghisellini}}},\ }\href {https://doi.org/10.1046/j.1365-8711.1998.01828.x}
  {\bibfield  {journal} {\bibinfo  {journal} {\mnras}\ }\textbf {\bibinfo
  {volume} {299}},\ \bibinfo {pages} {433} (\bibinfo {year} {1998})},\ \Eprint
  {https://arxiv.org/abs/astro-ph/9804103} {astro-ph/9804103} \BibitemShut
  {NoStop}%
\bibitem [{\citenamefont {{Maraschi}}\ \emph {et~al.}(1992)\citenamefont
  {{Maraschi}}, \citenamefont {{Ghisellini}},\ and\ \citenamefont
  {{Celotti}}}]{Maraschi1992}%
  \BibitemOpen
  \bibfield  {author} {\bibinfo {author} {\bibfnamefont {L.}~\bibnamefont
  {{Maraschi}}}, \bibinfo {author} {\bibfnamefont {G.}~\bibnamefont
  {{Ghisellini}}},\ and\ \bibinfo {author} {\bibfnamefont {A.}~\bibnamefont
  {{Celotti}}},\ }\href {https://doi.org/10.1086/186531} {\bibfield  {journal}
  {\bibinfo  {journal} {\apjl}\ }\textbf {\bibinfo {volume} {397}},\ \bibinfo
  {pages} {L5} (\bibinfo {year} {1992})}\BibitemShut {NoStop}%
\bibitem [{\citenamefont {{Dermer}}\ \emph {et~al.}(1992)\citenamefont
  {{Dermer}}, \citenamefont {{Schlickeiser}},\ and\ \citenamefont
  {{Mastichiadis}}}]{Dermer1992}%
  \BibitemOpen
  \bibfield  {author} {\bibinfo {author} {\bibfnamefont {C.~D.}\ \bibnamefont
  {{Dermer}}}, \bibinfo {author} {\bibfnamefont {R.}~\bibnamefont
  {{Schlickeiser}}},\ and\ \bibinfo {author} {\bibfnamefont {A.}~\bibnamefont
  {{Mastichiadis}}},\ }\href@noop {} {\bibfield  {journal} {\bibinfo  {journal}
  {\aap}\ }\textbf {\bibinfo {volume} {256}},\ \bibinfo {pages} {L27} (\bibinfo
  {year} {1992})}\BibitemShut {NoStop}%
\bibitem [{\citenamefont {{Sikora}}\ \emph {et~al.}(1994)\citenamefont
  {{Sikora}}, \citenamefont {{Begelman}},\ and\ \citenamefont
  {{Rees}}}]{Sikora1994}%
  \BibitemOpen
  \bibfield  {author} {\bibinfo {author} {\bibfnamefont {M.}~\bibnamefont
  {{Sikora}}}, \bibinfo {author} {\bibfnamefont {M.~C.}\ \bibnamefont
  {{Begelman}}},\ and\ \bibinfo {author} {\bibfnamefont {M.~J.}\ \bibnamefont
  {{Rees}}},\ }\href {https://doi.org/10.1086/173633} {\bibfield  {journal}
  {\bibinfo  {journal} {\apj}\ }\textbf {\bibinfo {volume} {421}},\ \bibinfo
  {pages} {153} (\bibinfo {year} {1994})}\BibitemShut {NoStop}%
\bibitem [{\citenamefont {{Bloom}}\ and\ \citenamefont
  {{Marscher}}(1996)}]{Bloom1996}%
  \BibitemOpen
  \bibfield  {author} {\bibinfo {author} {\bibfnamefont {S.~D.}\ \bibnamefont
  {{Bloom}}}\ and\ \bibinfo {author} {\bibfnamefont {A.~P.}\ \bibnamefont
  {{Marscher}}},\ }\href {https://doi.org/10.1086/177092} {\bibfield  {journal}
  {\bibinfo  {journal} {\apj}\ }\textbf {\bibinfo {volume} {461}},\ \bibinfo
  {pages} {657} (\bibinfo {year} {1996})}\BibitemShut {NoStop}%
\bibitem [{\citenamefont {{Blandford}}(2019)}]{2019arXiv190105164B}%
  \BibitemOpen
  \bibfield  {author} {\bibinfo {author} {\bibfnamefont {R.}~\bibnamefont
  {{Blandford}}},\ }\href@noop {} {\bibfield  {journal} {\bibinfo  {journal}
  {arXiv e-prints}\ ,\ \bibinfo {eid} {arXiv:1901.05164}} (\bibinfo {year}
  {2019})},\ \Eprint {https://arxiv.org/abs/1901.05164} {arXiv:1901.05164
  [astro-ph.HE]} \BibitemShut {NoStop}%
\bibitem [{\citenamefont {{Aharonian}}(2000)}]{aharonian00}%
  \BibitemOpen
  \bibfield  {author} {\bibinfo {author} {\bibfnamefont {F.~A.}\ \bibnamefont
  {{Aharonian}}},\ }\href {https://doi.org/10.1016/S1384-1076(00)00039-7}
  {\bibfield  {journal} {\bibinfo  {journal} {New Astron.}\ }\textbf {\bibinfo
  {volume} {5}},\ \bibinfo {pages} {377} (\bibinfo {year} {2000})},\ \Eprint
  {https://arxiv.org/abs/arXiv:astro-ph/0003159} {arXiv:astro-ph/0003159}
  \BibitemShut {NoStop}%
\bibitem [{\citenamefont {{M{\"u}cke}}\ and\ \citenamefont
  {{Protheroe}}(2001)}]{muecke01}%
  \BibitemOpen
  \bibfield  {author} {\bibinfo {author} {\bibfnamefont {A.}~\bibnamefont
  {{M{\"u}cke}}}\ and\ \bibinfo {author} {\bibfnamefont {R.~J.}\ \bibnamefont
  {{Protheroe}}},\ }\href {https://doi.org/10.1016/S0927-6505(00)00141-9}
  {\bibfield  {journal} {\bibinfo  {journal} {Astroparticle Physics}\ }\textbf
  {\bibinfo {volume} {15}},\ \bibinfo {pages} {121} (\bibinfo {year} {2001})},\
  \Eprint {https://arxiv.org/abs/astro-ph/0004052} {astro-ph/0004052}
  \BibitemShut {NoStop}%
\bibitem [{\citenamefont {{M{\"u}cke}}\ \emph {et~al.}(2003)\citenamefont
  {{M{\"u}cke}}, \citenamefont {{Protheroe}}, \citenamefont {{Engel}},
  \citenamefont {{Rachen}},\ and\ \citenamefont {{Stanev}}}]{muecke03}%
  \BibitemOpen
  \bibfield  {author} {\bibinfo {author} {\bibfnamefont {A.}~\bibnamefont
  {{M{\"u}cke}}}, \bibinfo {author} {\bibfnamefont {R.~J.}\ \bibnamefont
  {{Protheroe}}}, \bibinfo {author} {\bibfnamefont {R.}~\bibnamefont
  {{Engel}}}, \bibinfo {author} {\bibfnamefont {J.~P.}\ \bibnamefont
  {{Rachen}}},\ and\ \bibinfo {author} {\bibfnamefont {T.}~\bibnamefont
  {{Stanev}}},\ }\href {https://doi.org/10.1016/S0927-6505(02)00185-8}
  {\bibfield  {journal} {\bibinfo  {journal} {Astroparticle Physics}\ }\textbf
  {\bibinfo {volume} {18}},\ \bibinfo {pages} {593} (\bibinfo {year} {2003})},\
  \Eprint {https://arxiv.org/abs/arXiv:astro-ph/0206164}
  {arXiv:astro-ph/0206164} \BibitemShut {NoStop}%
\bibitem [{\citenamefont {{Petropoulou}}\ and\ \citenamefont
  {{Dimitrakoudis}}(2015)}]{petrodimi15}%
  \BibitemOpen
  \bibfield  {author} {\bibinfo {author} {\bibfnamefont {M.}~\bibnamefont
  {{Petropoulou}}}\ and\ \bibinfo {author} {\bibfnamefont {S.}~\bibnamefont
  {{Dimitrakoudis}}},\ }\href {https://doi.org/10.1093/mnras/stv1380}
  {\bibfield  {journal} {\bibinfo  {journal} {\mnras}\ }\textbf {\bibinfo
  {volume} {452}},\ \bibinfo {pages} {1303} (\bibinfo {year} {2015})},\ \Eprint
  {https://arxiv.org/abs/1506.05723} {arXiv:1506.05723 [astro-ph.HE]}
  \BibitemShut {NoStop}%
\bibitem [{\citenamefont {{Cerruti}}\ \emph {et~al.}(2015)\citenamefont
  {{Cerruti}}, \citenamefont {{Zech}}, \citenamefont {{Boisson}},\ and\
  \citenamefont {{Inoue}}}]{cerruti15}%
  \BibitemOpen
  \bibfield  {author} {\bibinfo {author} {\bibfnamefont {M.}~\bibnamefont
  {{Cerruti}}}, \bibinfo {author} {\bibfnamefont {A.}~\bibnamefont {{Zech}}},
  \bibinfo {author} {\bibfnamefont {C.}~\bibnamefont {{Boisson}}},\ and\
  \bibinfo {author} {\bibfnamefont {S.}~\bibnamefont {{Inoue}}},\ }\href
  {https://doi.org/10.1093/mnras/stu2691} {\bibfield  {journal} {\bibinfo
  {journal} {\mnras}\ }\textbf {\bibinfo {volume} {448}},\ \bibinfo {pages}
  {910} (\bibinfo {year} {2015})},\ \Eprint {https://arxiv.org/abs/1411.5968}
  {arXiv:1411.5968 [astro-ph.HE]} \BibitemShut {NoStop}%
\bibitem [{\citenamefont {{Mannheim}}(1993)}]{mannheim93}%
  \BibitemOpen
  \bibfield  {author} {\bibinfo {author} {\bibfnamefont {K.}~\bibnamefont
  {{Mannheim}}},\ }\href@noop {} {\bibfield  {journal} {\bibinfo  {journal}
  {\aap}\ }\textbf {\bibinfo {volume} {269}},\ \bibinfo {pages} {67} (\bibinfo
  {year} {1993})},\ \Eprint {https://arxiv.org/abs/astro-ph/9302006}
  {astro-ph/9302006} \BibitemShut {NoStop}%
\bibitem [{\citenamefont {Dermer}\ \emph {et~al.}(2012)\citenamefont {Dermer},
  \citenamefont {Murase},\ and\ \citenamefont {Takami}}]{Dermer:2012rg}%
  \BibitemOpen
  \bibfield  {author} {\bibinfo {author} {\bibfnamefont {C.~D.}\ \bibnamefont
  {Dermer}}, \bibinfo {author} {\bibfnamefont {K.}~\bibnamefont {Murase}},\
  and\ \bibinfo {author} {\bibfnamefont {H.}~\bibnamefont {Takami}},\ }\href
  {https://doi.org/10.1088/0004-637X/755/2/147} {\bibfield  {journal} {\bibinfo
   {journal} {Astrophys. J.}\ }\textbf {\bibinfo {volume} {755}},\ \bibinfo
  {pages} {147} (\bibinfo {year} {2012})},\ \Eprint
  {https://arxiv.org/abs/1203.6544} {arXiv:1203.6544 [astro-ph.HE]}
  \BibitemShut {NoStop}%
\bibitem [{\citenamefont {{Sahu}}\ \emph {et~al.}(2013)\citenamefont {{Sahu}},
  \citenamefont {{Oliveros}},\ and\ \citenamefont {{Sanabria}}}]{sahu13}%
  \BibitemOpen
  \bibfield  {author} {\bibinfo {author} {\bibfnamefont {S.}~\bibnamefont
  {{Sahu}}}, \bibinfo {author} {\bibfnamefont {A.~F.~O.}\ \bibnamefont
  {{Oliveros}}},\ and\ \bibinfo {author} {\bibfnamefont {J.~C.}\ \bibnamefont
  {{Sanabria}}},\ }\href {https://doi.org/10.1103/PhysRevD.87.103015}
  {\bibfield  {journal} {\bibinfo  {journal} {\prd}\ }\textbf {\bibinfo
  {volume} {87}},\ \bibinfo {eid} {103015} (\bibinfo {year} {2013})},\ \Eprint
  {https://arxiv.org/abs/1305.4985} {arXiv:1305.4985 [hep-ph]} \BibitemShut
  {NoStop}%
\bibitem [{\citenamefont {Petropoulou}\ \emph {et~al.}(2015)\citenamefont
  {Petropoulou}, \citenamefont {Dimitrakoudis}, \citenamefont {Padovani},
  \citenamefont {Mastichiadis},\ and\ \citenamefont
  {Resconi}}]{Petropoulou:2015upa}%
  \BibitemOpen
  \bibfield  {author} {\bibinfo {author} {\bibfnamefont {M.}~\bibnamefont
  {Petropoulou}}, \bibinfo {author} {\bibfnamefont {S.}~\bibnamefont
  {Dimitrakoudis}}, \bibinfo {author} {\bibfnamefont {P.}~\bibnamefont
  {Padovani}}, \bibinfo {author} {\bibfnamefont {A.}~\bibnamefont
  {Mastichiadis}},\ and\ \bibinfo {author} {\bibfnamefont {E.}~\bibnamefont
  {Resconi}},\ }\href {https://doi.org/10.1093/mnras/stv179} {\bibfield
  {journal} {\bibinfo  {journal} {Mon. Not. Roy. Astron. Soc.}\ }\textbf
  {\bibinfo {volume} {448}},\ \bibinfo {pages} {2412} (\bibinfo {year}
  {2015})},\ \Eprint {https://arxiv.org/abs/1501.07115} {arXiv:1501.07115
  [astro-ph.HE]} \BibitemShut {NoStop}%
\bibitem [{\citenamefont {{Petropoulou}}\ \emph {et~al.}(2016)\citenamefont
  {{Petropoulou}}, \citenamefont {{Coenders}},\ and\ \citenamefont
  {{Dimitrakoudis}}}]{Petropoulou2016}%
  \BibitemOpen
  \bibfield  {author} {\bibinfo {author} {\bibfnamefont {M.}~\bibnamefont
  {{Petropoulou}}}, \bibinfo {author} {\bibfnamefont {S.}~\bibnamefont
  {{Coenders}}},\ and\ \bibinfo {author} {\bibfnamefont {S.}~\bibnamefont
  {{Dimitrakoudis}}},\ }\href
  {https://doi.org/10.1016/j.astropartphys.2016.04.001} {\bibfield  {journal}
  {\bibinfo  {journal} {Astroparticle Physics}\ }\textbf {\bibinfo {volume}
  {80}},\ \bibinfo {pages} {115} (\bibinfo {year} {2016})},\ \Eprint
  {https://arxiv.org/abs/1603.06954} {arXiv:1603.06954 [astro-ph.HE]}
  \BibitemShut {NoStop}%
\bibitem [{\citenamefont {{Berezinskii}}\ and\ \citenamefont
  {{Smirnov}}(1975)}]{1975Ap&SS..32..461B}%
  \BibitemOpen
  \bibfield  {author} {\bibinfo {author} {\bibfnamefont {V.~S.}\ \bibnamefont
  {{Berezinskii}}}\ and\ \bibinfo {author} {\bibfnamefont {A.~I.}\ \bibnamefont
  {{Smirnov}}},\ }\href {https://doi.org/10.1007/BF00643157} {\bibfield
  {journal} {\bibinfo  {journal} {\apss}\ }\textbf {\bibinfo {volume} {32}},\
  \bibinfo {pages} {461} (\bibinfo {year} {1975})}\BibitemShut {NoStop}%
\bibitem [{\citenamefont {Essey}\ and\ \citenamefont
  {Kusenko}(2010)}]{Essey:2009zg}%
  \BibitemOpen
  \bibfield  {author} {\bibinfo {author} {\bibfnamefont {W.}~\bibnamefont
  {Essey}}\ and\ \bibinfo {author} {\bibfnamefont {A.}~\bibnamefont
  {Kusenko}},\ }\href {https://doi.org/10.1016/j.astropartphys.2009.11.007}
  {\bibfield  {journal} {\bibinfo  {journal} {Astropart. Phys.}\ }\textbf
  {\bibinfo {volume} {33}},\ \bibinfo {pages} {81} (\bibinfo {year} {2010})},\
  \Eprint {https://arxiv.org/abs/0905.1162} {arXiv:0905.1162 [astro-ph.HE]}
  \BibitemShut {NoStop}%
\bibitem [{\citenamefont {Essey}\ \emph {et~al.}(2010)\citenamefont {Essey},
  \citenamefont {Kalashev}, \citenamefont {Kusenko},\ and\ \citenamefont
  {Beacom}}]{Essey:2009ju}%
  \BibitemOpen
  \bibfield  {author} {\bibinfo {author} {\bibfnamefont {W.}~\bibnamefont
  {Essey}}, \bibinfo {author} {\bibfnamefont {O.~E.}\ \bibnamefont {Kalashev}},
  \bibinfo {author} {\bibfnamefont {A.}~\bibnamefont {Kusenko}},\ and\ \bibinfo
  {author} {\bibfnamefont {J.~F.}\ \bibnamefont {Beacom}},\ }\href
  {https://doi.org/10.1103/PhysRevLett.104.141102} {\bibfield  {journal}
  {\bibinfo  {journal} {Phys. Rev. Lett.}\ }\textbf {\bibinfo {volume} {104}},\
  \bibinfo {pages} {141102} (\bibinfo {year} {2010})},\ \Eprint
  {https://arxiv.org/abs/0912.3976} {arXiv:0912.3976 [astro-ph.HE]}
  \BibitemShut {NoStop}%
\bibitem [{\citenamefont {Murase}\ \emph {et~al.}(2012)\citenamefont {Murase},
  \citenamefont {Dermer}, \citenamefont {Takami},\ and\ \citenamefont
  {Migliori}}]{Murase:2011cy}%
  \BibitemOpen
  \bibfield  {author} {\bibinfo {author} {\bibfnamefont {K.}~\bibnamefont
  {Murase}}, \bibinfo {author} {\bibfnamefont {C.~D.}\ \bibnamefont {Dermer}},
  \bibinfo {author} {\bibfnamefont {H.}~\bibnamefont {Takami}},\ and\ \bibinfo
  {author} {\bibfnamefont {G.}~\bibnamefont {Migliori}},\ }\href
  {https://doi.org/10.1088/0004-637X/749/1/63} {\bibfield  {journal} {\bibinfo
  {journal} {Astrophys. J.}\ }\textbf {\bibinfo {volume} {749}},\ \bibinfo
  {pages} {63} (\bibinfo {year} {2012})},\ \Eprint
  {https://arxiv.org/abs/1107.5576} {arXiv:1107.5576 [astro-ph.HE]}
  \BibitemShut {NoStop}%
\bibitem [{\citenamefont {Oikonomou}\ \emph {et~al.}(2014)\citenamefont
  {Oikonomou}, \citenamefont {Murase},\ and\ \citenamefont
  {Kotera}}]{Oikonomou:2014xea}%
  \BibitemOpen
  \bibfield  {author} {\bibinfo {author} {\bibfnamefont {F.}~\bibnamefont
  {Oikonomou}}, \bibinfo {author} {\bibfnamefont {K.}~\bibnamefont {Murase}},\
  and\ \bibinfo {author} {\bibfnamefont {K.}~\bibnamefont {Kotera}},\ }\href
  {https://doi.org/10.1051/0004-6361/201423798} {\bibfield  {journal} {\bibinfo
   {journal} {Astron. Astrophys.}\ }\textbf {\bibinfo {volume} {568}},\
  \bibinfo {pages} {A110} (\bibinfo {year} {2014})},\ \Eprint
  {https://arxiv.org/abs/1406.6075} {arXiv:1406.6075 [astro-ph.HE]}
  \BibitemShut {NoStop}%
\bibitem [{\citenamefont {{Hillas}}(1984)}]{Hillas1984}%
  \BibitemOpen
  \bibfield  {author} {\bibinfo {author} {\bibfnamefont {A.~M.}\ \bibnamefont
  {{Hillas}}},\ }\href {https://doi.org/10.1146/annurev.aa.22.090184.002233}
  {\bibfield  {journal} {\bibinfo  {journal} {Annual Review of Astronomy and
  Astrophysics}\ }\textbf {\bibinfo {volume} {22}},\ \bibinfo {pages} {425}
  (\bibinfo {year} {1984})}\BibitemShut {NoStop}%
\bibitem [{\citenamefont {{Blandford}}\ \emph {et~al.}(2018)\citenamefont
  {{Blandford}}, \citenamefont {{Meier}},\ and\ \citenamefont
  {{Readhead}}}]{2018arXiv181206025B}%
  \BibitemOpen
  \bibfield  {author} {\bibinfo {author} {\bibfnamefont {R.}~\bibnamefont
  {{Blandford}}}, \bibinfo {author} {\bibfnamefont {D.}~\bibnamefont
  {{Meier}}},\ and\ \bibinfo {author} {\bibfnamefont {A.}~\bibnamefont
  {{Readhead}}},\ }\href@noop {} {\bibfield  {journal} {\bibinfo  {journal}
  {arXiv e-prints}\ ,\ \bibinfo {eid} {arXiv:1812.06025}} (\bibinfo {year}
  {2018})},\ \Eprint {https://arxiv.org/abs/1812.06025} {arXiv:1812.06025
  [astro-ph.HE]} \BibitemShut {NoStop}%
\bibitem [{\citenamefont {Biermann}\ and\ \citenamefont
  {Strittmatter}(1987)}]{Biermann:1987ep}%
  \BibitemOpen
  \bibfield  {author} {\bibinfo {author} {\bibfnamefont {P.~L.}\ \bibnamefont
  {Biermann}}\ and\ \bibinfo {author} {\bibfnamefont {P.~A.}\ \bibnamefont
  {Strittmatter}},\ }\href {https://doi.org/10.1086/165759} {\bibfield
  {journal} {\bibinfo  {journal} {Astrophys. J.}\ }\textbf {\bibinfo {volume}
  {322}},\ \bibinfo {pages} {643} (\bibinfo {year} {1987})}\BibitemShut
  {NoStop}%
\bibitem [{\citenamefont {Ostrowski}(1998)}]{Ostrowski:1998ic}%
  \BibitemOpen
  \bibfield  {author} {\bibinfo {author} {\bibfnamefont {M.}~\bibnamefont
  {Ostrowski}},\ }\href@noop {} {\bibfield  {journal} {\bibinfo  {journal}
  {Astron. Astrophys.}\ }\textbf {\bibinfo {volume} {335}},\ \bibinfo {pages}
  {134} (\bibinfo {year} {1998})},\ \Eprint
  {https://arxiv.org/abs/astro-ph/9803299} {arXiv:astro-ph/9803299 [astro-ph]}
  \BibitemShut {NoStop}%
\bibitem [{\citenamefont {Rieger}\ and\ \citenamefont
  {Duffy}(2004)}]{Rieger:2004jz}%
  \BibitemOpen
  \bibfield  {author} {\bibinfo {author} {\bibfnamefont {F.~M.}\ \bibnamefont
  {Rieger}}\ and\ \bibinfo {author} {\bibfnamefont {P.}~\bibnamefont {Duffy}},\
  }\href {https://doi.org/10.1086/425167} {\bibfield  {journal} {\bibinfo
  {journal} {Astrophys. J.}\ }\textbf {\bibinfo {volume} {617}},\ \bibinfo
  {pages} {155} (\bibinfo {year} {2004})},\ \Eprint
  {https://arxiv.org/abs/astro-ph/0410269} {arXiv:astro-ph/0410269 [astro-ph]}
  \BibitemShut {NoStop}%
\bibitem [{\citenamefont {Pe'er}\ \emph {et~al.}(2009)\citenamefont {Pe'er},
  \citenamefont {Murase},\ and\ \citenamefont {Meszaros}}]{Peer:2009vnw}%
  \BibitemOpen
  \bibfield  {author} {\bibinfo {author} {\bibfnamefont {A.}~\bibnamefont
  {Pe'er}}, \bibinfo {author} {\bibfnamefont {K.}~\bibnamefont {Murase}},\ and\
  \bibinfo {author} {\bibfnamefont {P.}~\bibnamefont {Meszaros}},\ }\href
  {https://doi.org/10.1103/PhysRevD.80.123018} {\bibfield  {journal} {\bibinfo
  {journal} {Phys. Rev.}\ }\textbf {\bibinfo {volume} {D80}},\ \bibinfo {pages}
  {123018} (\bibinfo {year} {2009})},\ \Eprint
  {https://arxiv.org/abs/0911.1776} {arXiv:0911.1776 [astro-ph.HE]}
  \BibitemShut {NoStop}%
\bibitem [{\citenamefont {Kimura}\ \emph {et~al.}(2018)\citenamefont {Kimura},
  \citenamefont {Murase},\ and\ \citenamefont {Zhang}}]{Kimura:2017ubz}%
  \BibitemOpen
  \bibfield  {author} {\bibinfo {author} {\bibfnamefont {S.~S.}\ \bibnamefont
  {Kimura}}, \bibinfo {author} {\bibfnamefont {K.}~\bibnamefont {Murase}},\
  and\ \bibinfo {author} {\bibfnamefont {B.~T.}\ \bibnamefont {Zhang}},\ }\href
  {https://doi.org/10.1103/PhysRevD.97.023026} {\bibfield  {journal} {\bibinfo
  {journal} {Phys. Rev.}\ }\textbf {\bibinfo {volume} {D97}},\ \bibinfo {pages}
  {023026} (\bibinfo {year} {2018})},\ \Eprint
  {https://arxiv.org/abs/1705.05027} {arXiv:1705.05027 [astro-ph.HE]}
  \BibitemShut {NoStop}%
\bibitem [{\citenamefont {Caprioli}(2015)}]{Caprioli:2015zka}%
  \BibitemOpen
  \bibfield  {author} {\bibinfo {author} {\bibfnamefont {D.}~\bibnamefont
  {Caprioli}},\ }\href {https://doi.org/10.1088/2041-8205/811/2/L38} {\bibfield
   {journal} {\bibinfo  {journal} {Astrophys. J.}\ }\textbf {\bibinfo {volume}
  {811}},\ \bibinfo {pages} {L38} (\bibinfo {year} {2015})},\ \Eprint
  {https://arxiv.org/abs/1505.06739} {arXiv:1505.06739 [astro-ph.HE]}
  \BibitemShut {NoStop}%
\bibitem [{\citenamefont {{Takahara}}(1990)}]{1990PThPh..83.1071T}%
  \BibitemOpen
  \bibfield  {author} {\bibinfo {author} {\bibfnamefont {F.}~\bibnamefont
  {{Takahara}}},\ }\href {https://doi.org/10.1143/PTP.83.1071} {\bibfield
  {journal} {\bibinfo  {journal} {Progress of Theoretical Physics}\ }\textbf
  {\bibinfo {volume} {83}},\ \bibinfo {pages} {1071} (\bibinfo {year}
  {1990})}\BibitemShut {NoStop}%
\bibitem [{\citenamefont {Rachen}\ and\ \citenamefont
  {Biermann}(1993)}]{Rachen:1992pg}%
  \BibitemOpen
  \bibfield  {author} {\bibinfo {author} {\bibfnamefont {J.~P.}\ \bibnamefont
  {Rachen}}\ and\ \bibinfo {author} {\bibfnamefont {P.~L.}\ \bibnamefont
  {Biermann}},\ }\href@noop {} {\bibfield  {journal} {\bibinfo  {journal}
  {Astron. Astrophys.}\ }\textbf {\bibinfo {volume} {272}},\ \bibinfo {pages}
  {161} (\bibinfo {year} {1993})},\ \Eprint
  {https://arxiv.org/abs/astro-ph/9301010} {arXiv:astro-ph/9301010 [astro-ph]}
  \BibitemShut {NoStop}%
\bibitem [{\citenamefont {Berezhko}(2008)}]{Berezhko:2008xn}%
  \BibitemOpen
  \bibfield  {author} {\bibinfo {author} {\bibfnamefont {E.~G.}\ \bibnamefont
  {Berezhko}},\ }\href {https://doi.org/10.1086/592233} {\bibfield  {journal}
  {\bibinfo  {journal} {Astrophys. J.}\ }\textbf {\bibinfo {volume} {684}},\
  \bibinfo {pages} {L69} (\bibinfo {year} {2008})},\ \Eprint
  {https://arxiv.org/abs/0809.0734} {arXiv:0809.0734 [astro-ph]} \BibitemShut
  {NoStop}%
\bibitem [{\citenamefont {{Stecker}}\ \emph {et~al.}(1991)\citenamefont
  {{Stecker}}, \citenamefont {{Done}}, \citenamefont {{Salamon}},\ and\
  \citenamefont {{Sommers}}}]{Stecker1991}%
  \BibitemOpen
  \bibfield  {author} {\bibinfo {author} {\bibfnamefont {F.~W.}\ \bibnamefont
  {{Stecker}}}, \bibinfo {author} {\bibfnamefont {C.}~\bibnamefont {{Done}}},
  \bibinfo {author} {\bibfnamefont {M.~H.}\ \bibnamefont {{Salamon}}},\ and\
  \bibinfo {author} {\bibfnamefont {P.}~\bibnamefont {{Sommers}}},\ }\href
  {https://doi.org/10.1103/PhysRevLett.66.2697} {\bibfield  {journal} {\bibinfo
   {journal} {Physical Review Letters}\ }\textbf {\bibinfo {volume} {66}},\
  \bibinfo {pages} {2697} (\bibinfo {year} {1991})}\BibitemShut {NoStop}%
\bibitem [{\citenamefont {Alvarez-Muniz}\ and\ \citenamefont
  {Meszaros}(2004)}]{AlvarezMuniz:2004uz}%
  \BibitemOpen
  \bibfield  {author} {\bibinfo {author} {\bibfnamefont {J.}~\bibnamefont
  {Alvarez-Muniz}}\ and\ \bibinfo {author} {\bibfnamefont {P.}~\bibnamefont
  {Meszaros}},\ }\href {https://doi.org/10.1103/PhysRevD.70.123001} {\bibfield
  {journal} {\bibinfo  {journal} {Phys. Rev.}\ }\textbf {\bibinfo {volume}
  {D70}},\ \bibinfo {pages} {123001} (\bibinfo {year} {2004})},\ \Eprint
  {https://arxiv.org/abs/astro-ph/0409034} {arXiv:astro-ph/0409034 [astro-ph]}
  \BibitemShut {NoStop}%
\bibitem [{\citenamefont {Stecker}(2005)}]{Stecker:2005hn}%
  \BibitemOpen
  \bibfield  {author} {\bibinfo {author} {\bibfnamefont {F.~W.}\ \bibnamefont
  {Stecker}},\ }\href {https://doi.org/10.1103/PhysRevD.72.107301} {\bibfield
  {journal} {\bibinfo  {journal} {Phys. Rev.}\ }\textbf {\bibinfo {volume}
  {D72}},\ \bibinfo {pages} {107301} (\bibinfo {year} {2005})},\ \Eprint
  {https://arxiv.org/abs/astro-ph/0510537} {arXiv:astro-ph/0510537 [astro-ph]}
  \BibitemShut {NoStop}%
\bibitem [{\citenamefont {Stecker}(2013)}]{Stecker:2013fxa}%
  \BibitemOpen
  \bibfield  {author} {\bibinfo {author} {\bibfnamefont {F.~W.}\ \bibnamefont
  {Stecker}},\ }\href {https://doi.org/10.1103/PhysRevD.88.047301} {\bibfield
  {journal} {\bibinfo  {journal} {Phys. Rev.}\ }\textbf {\bibinfo {volume}
  {D88}},\ \bibinfo {pages} {047301} (\bibinfo {year} {2013})},\ \Eprint
  {https://arxiv.org/abs/1305.7404} {arXiv:1305.7404 [astro-ph.HE]}
  \BibitemShut {NoStop}%
\bibitem [{\citenamefont {Becker~Tjus}\ \emph {et~al.}(2014)\citenamefont
  {Becker~Tjus}, \citenamefont {Eichmann}, \citenamefont {Halzen},
  \citenamefont {Kheirandish},\ and\ \citenamefont {Saba}}]{Tjus:2014dna}%
  \BibitemOpen
  \bibfield  {author} {\bibinfo {author} {\bibfnamefont {J.}~\bibnamefont
  {Becker~Tjus}}, \bibinfo {author} {\bibfnamefont {B.}~\bibnamefont
  {Eichmann}}, \bibinfo {author} {\bibfnamefont {F.}~\bibnamefont {Halzen}},
  \bibinfo {author} {\bibfnamefont {A.}~\bibnamefont {Kheirandish}},\ and\
  \bibinfo {author} {\bibfnamefont {S.~M.}\ \bibnamefont {Saba}},\ }\href
  {https://doi.org/10.1103/PhysRevD.89.123005} {\bibfield  {journal} {\bibinfo
  {journal} {Phys. Rev.}\ }\textbf {\bibinfo {volume} {D89}},\ \bibinfo {pages}
  {123005} (\bibinfo {year} {2014})},\ \Eprint
  {https://arxiv.org/abs/1406.0506} {arXiv:1406.0506 [astro-ph.HE]}
  \BibitemShut {NoStop}%
\bibitem [{\citenamefont {Kimura}\ \emph {et~al.}(2015)\citenamefont {Kimura},
  \citenamefont {Murase},\ and\ \citenamefont {Toma}}]{Kimura:2014jba}%
  \BibitemOpen
  \bibfield  {author} {\bibinfo {author} {\bibfnamefont {S.~S.}\ \bibnamefont
  {Kimura}}, \bibinfo {author} {\bibfnamefont {K.}~\bibnamefont {Murase}},\
  and\ \bibinfo {author} {\bibfnamefont {K.}~\bibnamefont {Toma}},\ }\href
  {https://doi.org/10.1088/0004-637X/806/2/159} {\bibfield  {journal} {\bibinfo
   {journal} {Astrophys. J.}\ }\textbf {\bibinfo {volume} {806}},\ \bibinfo
  {pages} {159} (\bibinfo {year} {2015})},\ \Eprint
  {https://arxiv.org/abs/1411.3588} {arXiv:1411.3588 [astro-ph.HE]}
  \BibitemShut {NoStop}%
\bibitem [{\citenamefont {{Mannheim}}\ \emph {et~al.}(1992)\citenamefont
  {{Mannheim}}, \citenamefont {{Stanev}},\ and\ \citenamefont
  {{Biermann}}}]{1992A&A...260L...1M}%
  \BibitemOpen
  \bibfield  {author} {\bibinfo {author} {\bibfnamefont {K.}~\bibnamefont
  {{Mannheim}}}, \bibinfo {author} {\bibfnamefont {T.}~\bibnamefont
  {{Stanev}}},\ and\ \bibinfo {author} {\bibfnamefont {P.~L.}\ \bibnamefont
  {{Biermann}}},\ }\href@noop {} {\bibfield  {journal} {\bibinfo  {journal}
  {\aap}\ }\textbf {\bibinfo {volume} {260}},\ \bibinfo {pages} {L1} (\bibinfo
  {year} {1992})}\BibitemShut {NoStop}%
\bibitem [{\citenamefont {{Atoyan}}\ and\ \citenamefont
  {{Dermer}}(2001)}]{Atoyan2001}%
  \BibitemOpen
  \bibfield  {author} {\bibinfo {author} {\bibfnamefont {A.}~\bibnamefont
  {{Atoyan}}}\ and\ \bibinfo {author} {\bibfnamefont {C.~D.}\ \bibnamefont
  {{Dermer}}},\ }\href {https://doi.org/10.1103/PhysRevLett.87.221102}
  {\bibfield  {journal} {\bibinfo  {journal} {\prl}\ }\textbf {\bibinfo
  {volume} {87}},\ \bibinfo {eid} {221102} (\bibinfo {year} {2001})},\ \Eprint
  {https://arxiv.org/abs/astro-ph/0108053} {arXiv:astro-ph/0108053 [astro-ph]}
  \BibitemShut {NoStop}%
\bibitem [{\citenamefont {{Atoyan}}\ and\ \citenamefont
  {{Dermer}}(2004)}]{Atoyan2004}%
  \BibitemOpen
  \bibfield  {author} {\bibinfo {author} {\bibfnamefont {A.}~\bibnamefont
  {{Atoyan}}}\ and\ \bibinfo {author} {\bibfnamefont {C.~D.}\ \bibnamefont
  {{Dermer}}},\ }\href {https://doi.org/10.1016/j.newar.2003.12.046} {\bibfield
   {journal} {\bibinfo  {journal} {New Astronomy Reviews}\ }\textbf {\bibinfo
  {volume} {48}},\ \bibinfo {pages} {381} (\bibinfo {year} {2004})},\ \Eprint
  {https://arxiv.org/abs/astro-ph/0402646} {arXiv:astro-ph/0402646 [astro-ph]}
  \BibitemShut {NoStop}%
\bibitem [{\citenamefont {{Dimitrakoudis}}\ \emph {et~al.}(2014)\citenamefont
  {{Dimitrakoudis}}, \citenamefont {{Petropoulou}},\ and\ \citenamefont
  {{Mastichiadis}}}]{DPM14}%
  \BibitemOpen
  \bibfield  {author} {\bibinfo {author} {\bibfnamefont {S.}~\bibnamefont
  {{Dimitrakoudis}}}, \bibinfo {author} {\bibfnamefont {M.}~\bibnamefont
  {{Petropoulou}}},\ and\ \bibinfo {author} {\bibfnamefont {A.}~\bibnamefont
  {{Mastichiadis}}},\ }\href
  {https://doi.org/10.1016/j.astropartphys.2013.10.005} {\bibfield  {journal}
  {\bibinfo  {journal} {Astroparticle Physics}\ }\textbf {\bibinfo {volume}
  {54}},\ \bibinfo {pages} {61} (\bibinfo {year} {2014})},\ \Eprint
  {https://arxiv.org/abs/1310.7923} {arXiv:1310.7923 [astro-ph.HE]}
  \BibitemShut {NoStop}%
\bibitem [{\citenamefont {Murase}\ \emph {et~al.}(2014)\citenamefont {Murase},
  \citenamefont {Inoue},\ and\ \citenamefont {Dermer}}]{Murase:2014foa}%
  \BibitemOpen
  \bibfield  {author} {\bibinfo {author} {\bibfnamefont {K.}~\bibnamefont
  {Murase}}, \bibinfo {author} {\bibfnamefont {Y.}~\bibnamefont {Inoue}},\ and\
  \bibinfo {author} {\bibfnamefont {C.~D.}\ \bibnamefont {Dermer}},\ }\href
  {https://doi.org/10.1103/PhysRevD.90.023007} {\bibfield  {journal} {\bibinfo
  {journal} {Phys. Rev.}\ }\textbf {\bibinfo {volume} {D90}},\ \bibinfo {pages}
  {023007} (\bibinfo {year} {2014})},\ \Eprint
  {https://arxiv.org/abs/1403.4089} {arXiv:1403.4089 [astro-ph.HE]}
  \BibitemShut {NoStop}%
\bibitem [{\citenamefont {Dermer}\ \emph {et~al.}(2014)\citenamefont {Dermer},
  \citenamefont {Murase},\ and\ \citenamefont {Inoue}}]{Dermer:2014vaa}%
  \BibitemOpen
  \bibfield  {author} {\bibinfo {author} {\bibfnamefont {C.~D.}\ \bibnamefont
  {Dermer}}, \bibinfo {author} {\bibfnamefont {K.}~\bibnamefont {Murase}},\
  and\ \bibinfo {author} {\bibfnamefont {Y.}~\bibnamefont {Inoue}},\ }\href
  {https://doi.org/10.1016/j.jheap.2014.09.001} {\bibfield  {journal} {\bibinfo
   {journal} {JHEAp}\ }\textbf {\bibinfo {volume} {3-4}},\ \bibinfo {pages}
  {29} (\bibinfo {year} {2014})},\ \Eprint {https://arxiv.org/abs/1406.2633}
  {arXiv:1406.2633 [astro-ph.HE]} \BibitemShut {NoStop}%
\bibitem [{\citenamefont {Tamborra}\ \emph {et~al.}(2014)\citenamefont
  {Tamborra}, \citenamefont {Ando},\ and\ \citenamefont
  {Murase}}]{Tamborra:2014xia}%
  \BibitemOpen
  \bibfield  {author} {\bibinfo {author} {\bibfnamefont {I.}~\bibnamefont
  {Tamborra}}, \bibinfo {author} {\bibfnamefont {S.}~\bibnamefont {Ando}},\
  and\ \bibinfo {author} {\bibfnamefont {K.}~\bibnamefont {Murase}},\ }\href
  {https://doi.org/10.1088/1475-7516/2014/09/043} {\bibfield  {journal}
  {\bibinfo  {journal} {JCAP}\ }\textbf {\bibinfo {volume} {1409}},\ \bibinfo
  {pages} {043}},\ \Eprint {https://arxiv.org/abs/1404.1189} {arXiv:1404.1189
  [astro-ph.HE]} \BibitemShut {NoStop}%
\bibitem [{\citenamefont {Hooper}(2016)}]{Hooper:2016jls}%
  \BibitemOpen
  \bibfield  {author} {\bibinfo {author} {\bibfnamefont {D.}~\bibnamefont
  {Hooper}},\ }\href {https://doi.org/10.1088/1475-7516/2016/09/002} {\bibfield
   {journal} {\bibinfo  {journal} {JCAP}\ }\textbf {\bibinfo {volume}
  {1609}}\bibfield  {number} {\bibinfo  {number} { (09)},\ \bibinfo {pages}
  {002}},\ }\Eprint {https://arxiv.org/abs/1605.06504} {arXiv:1605.06504
  [astro-ph.HE]} \BibitemShut {NoStop}%
\bibitem [{\citenamefont {Wang}\ and\ \citenamefont
  {Loeb}(2016)}]{Wang:2016vbf}%
  \BibitemOpen
  \bibfield  {author} {\bibinfo {author} {\bibfnamefont {X.}~\bibnamefont
  {Wang}}\ and\ \bibinfo {author} {\bibfnamefont {A.}~\bibnamefont {Loeb}},\
  }\href {https://doi.org/10.1088/1475-7516/2016/12/012} {\bibfield  {journal}
  {\bibinfo  {journal} {JCAP}\ }\textbf {\bibinfo {volume} {1612}}\bibfield
  {number} {\bibinfo  {number} { (12)},\ \bibinfo {pages} {012}},\ }\Eprint
  {https://arxiv.org/abs/1607.06476} {arXiv:1607.06476 [astro-ph.HE]}
  \BibitemShut {NoStop}%
\bibitem [{\citenamefont {Lamastra}\ \emph {et~al.}(2017)\citenamefont
  {Lamastra}, \citenamefont {Menci}, \citenamefont {Fiore}, \citenamefont
  {Antonelli}, \citenamefont {Colafrancesco}, \citenamefont {Guetta},\ and\
  \citenamefont {Stamerra}}]{Lamastra:2017iyo}%
  \BibitemOpen
  \bibfield  {author} {\bibinfo {author} {\bibfnamefont {A.}~\bibnamefont
  {Lamastra}}, \bibinfo {author} {\bibfnamefont {N.}~\bibnamefont {Menci}},
  \bibinfo {author} {\bibfnamefont {F.}~\bibnamefont {Fiore}}, \bibinfo
  {author} {\bibfnamefont {L.~A.}\ \bibnamefont {Antonelli}}, \bibinfo {author}
  {\bibfnamefont {S.}~\bibnamefont {Colafrancesco}}, \bibinfo {author}
  {\bibfnamefont {D.}~\bibnamefont {Guetta}},\ and\ \bibinfo {author}
  {\bibfnamefont {A.}~\bibnamefont {Stamerra}},\ }\href
  {https://doi.org/10.1051/0004-6361/201731452} {\bibfield  {journal} {\bibinfo
   {journal} {Astron. Astrophys.}\ }\textbf {\bibinfo {volume} {607}},\
  \bibinfo {pages} {A18} (\bibinfo {year} {2017})},\ \Eprint
  {https://arxiv.org/abs/1709.03497} {arXiv:1709.03497 [astro-ph.HE]}
  \BibitemShut {NoStop}%
\bibitem [{\citenamefont {Liu}\ \emph {et~al.}(2018)\citenamefont {Liu},
  \citenamefont {Murase}, \citenamefont {Inoue}, \citenamefont {Ge},\ and\
  \citenamefont {Wang}}]{Liu:2017bjr}%
  \BibitemOpen
  \bibfield  {author} {\bibinfo {author} {\bibfnamefont {R.-Y.}\ \bibnamefont
  {Liu}}, \bibinfo {author} {\bibfnamefont {K.}~\bibnamefont {Murase}},
  \bibinfo {author} {\bibfnamefont {S.}~\bibnamefont {Inoue}}, \bibinfo
  {author} {\bibfnamefont {C.}~\bibnamefont {Ge}},\ and\ \bibinfo {author}
  {\bibfnamefont {X.-Y.}\ \bibnamefont {Wang}},\ }\href
  {https://doi.org/10.3847/1538-4357/aaba74} {\bibfield  {journal} {\bibinfo
  {journal} {Astrophys. J.}\ }\textbf {\bibinfo {volume} {858}},\ \bibinfo
  {pages} {9} (\bibinfo {year} {2018})},\ \Eprint
  {https://arxiv.org/abs/1712.10168} {arXiv:1712.10168 [astro-ph.HE]}
  \BibitemShut {NoStop}%
\bibitem [{\citenamefont {Hooper}\ \emph {et~al.}(2019)\citenamefont {Hooper},
  \citenamefont {Linden},\ and\ \citenamefont {Vieregg}}]{Hooper:2018wyk}%
  \BibitemOpen
  \bibfield  {author} {\bibinfo {author} {\bibfnamefont {D.}~\bibnamefont
  {Hooper}}, \bibinfo {author} {\bibfnamefont {T.}~\bibnamefont {Linden}},\
  and\ \bibinfo {author} {\bibfnamefont {A.}~\bibnamefont {Vieregg}},\ }\href
  {https://doi.org/10.1088/1475-7516/2019/02/012} {\bibfield  {journal}
  {\bibinfo  {journal} {JCAP}\ }\textbf {\bibinfo {volume} {1902}},\ \bibinfo
  {pages} {012}},\ \Eprint {https://arxiv.org/abs/1810.02823} {arXiv:1810.02823
  [astro-ph.HE]} \BibitemShut {NoStop}%
\bibitem [{\citenamefont {Murase}\ \emph {et~al.}(2018)\citenamefont {Murase},
  \citenamefont {Oikonomou},\ and\ \citenamefont
  {Petropoulou}}]{Murase:2018iyl}%
  \BibitemOpen
  \bibfield  {author} {\bibinfo {author} {\bibfnamefont {K.}~\bibnamefont
  {Murase}}, \bibinfo {author} {\bibfnamefont {F.}~\bibnamefont {Oikonomou}},\
  and\ \bibinfo {author} {\bibfnamefont {M.}~\bibnamefont {Petropoulou}},\
  }\href {https://doi.org/10.3847/1538-4357/aada00} {\bibfield  {journal}
  {\bibinfo  {journal} {Astrophys. J.}\ }\textbf {\bibinfo {volume} {865}},\
  \bibinfo {pages} {124} (\bibinfo {year} {2018})},\ \Eprint
  {https://arxiv.org/abs/1807.04748} {arXiv:1807.04748 [astro-ph.HE]}
  \BibitemShut {NoStop}%
\bibitem [{\citenamefont {Kotera}\ \emph {et~al.}(2009)\citenamefont {Kotera},
  \citenamefont {Allard}, \citenamefont {Murase}, \citenamefont {Aoi},
  \citenamefont {Dubois}, \citenamefont {Pierog},\ and\ \citenamefont
  {Nagataki}}]{Kotera:2009ms}%
  \BibitemOpen
  \bibfield  {author} {\bibinfo {author} {\bibfnamefont {K.}~\bibnamefont
  {Kotera}}, \bibinfo {author} {\bibfnamefont {D.}~\bibnamefont {Allard}},
  \bibinfo {author} {\bibfnamefont {K.}~\bibnamefont {Murase}}, \bibinfo
  {author} {\bibfnamefont {J.}~\bibnamefont {Aoi}}, \bibinfo {author}
  {\bibfnamefont {Y.}~\bibnamefont {Dubois}}, \bibinfo {author} {\bibfnamefont
  {T.}~\bibnamefont {Pierog}},\ and\ \bibinfo {author} {\bibfnamefont
  {S.}~\bibnamefont {Nagataki}},\ }\href
  {https://doi.org/10.1088/0004-637X/707/1/370} {\bibfield  {journal} {\bibinfo
   {journal} {Astrophys. J.}\ }\textbf {\bibinfo {volume} {707}},\ \bibinfo
  {pages} {370} (\bibinfo {year} {2009})},\ \Eprint
  {https://arxiv.org/abs/0907.2433} {arXiv:0907.2433 [astro-ph.HE]}
  \BibitemShut {NoStop}%
\bibitem [{\citenamefont {{Fang}}\ and\ \citenamefont
  {{Murase}}(2018)}]{2018NatPh..14..396F}%
  \BibitemOpen
  \bibfield  {author} {\bibinfo {author} {\bibfnamefont {K.}~\bibnamefont
  {{Fang}}}\ and\ \bibinfo {author} {\bibfnamefont {K.}~\bibnamefont
  {{Murase}}},\ }\href {https://doi.org/10.1038/s41567-017-0025-4} {\bibfield
  {journal} {\bibinfo  {journal} {Nature Physics}\ }\textbf {\bibinfo {volume}
  {14}},\ \bibinfo {pages} {396} (\bibinfo {year} {2018})},\ \Eprint
  {https://arxiv.org/abs/1704.00015} {arXiv:1704.00015 [astro-ph.HE]}
  \BibitemShut {NoStop}%
\bibitem [{\citenamefont {Aartsen}\ \emph
  {et~al.}(2018{\natexlab{a}})\citenamefont {Aartsen} \emph
  {et~al.}}]{IceCube:2018dnn}%
  \BibitemOpen
  \bibfield  {author} {\bibinfo {author} {\bibfnamefont {M.~G.}\ \bibnamefont
  {Aartsen}} \emph {et~al.} (\bibinfo {collaboration} {IceCube, Fermi-LAT,
  MAGIC, AGILE, ASAS-SN, HAWC, H.E.S.S., INTEGRAL, Kanata, Kiso, Kapteyn,
  Liverpool Telescope, Subaru, Swift NuSTAR, VERITAS, VLA/17B-403}),\ }\href
  {https://doi.org/10.1126/science.aat1378} {\bibfield  {journal} {\bibinfo
  {journal} {Science}\ }\textbf {\bibinfo {volume} {361}},\ \bibinfo {pages}
  {eaat1378} (\bibinfo {year} {2018}{\natexlab{a}})},\ \Eprint
  {https://arxiv.org/abs/1807.08816} {arXiv:1807.08816 [astro-ph.HE]}
  \BibitemShut {NoStop}%
\bibitem [{LAT()}]{LATSensitivity}%
  \BibitemOpen
  \href@noop {} {\bibinfo {title} {{LAT P8R2 Performance}}},\ \bibinfo
  {howpublished}
  {\url{http://www.slac.stanford.edu/exp/glast/groups/canda/lat_Performance.htm}},\
  \bibinfo {note} {accessed: 2019-03-07}\BibitemShut {NoStop}%
\bibitem [{\citenamefont {{Park}}(2015)}]{2015ICRC...34..771P}%
  \BibitemOpen
  \bibfield  {author} {\bibinfo {author} {\bibfnamefont {N.}~\bibnamefont
  {{Park}}} (\bibinfo {collaboration} {VERITAS}),\ }in\ \href@noop {} {\emph
  {\bibinfo {booktitle} {34th International Cosmic Ray Conference
  (ICRC2015)}}},\ Vol.~\bibinfo {volume} {34}\ (\bibinfo {year} {2015})\ p.\
  \bibinfo {pages} {771},\ \Eprint {https://arxiv.org/abs/1508.07070}
  {arXiv:1508.07070 [astro-ph.IM]} \BibitemShut {NoStop}%
\bibitem [{\citenamefont {Abeysekara}\ \emph
  {et~al.}(2017{\natexlab{a}})\citenamefont {Abeysekara} \emph
  {et~al.}}]{Abeysekara:2017mjj}%
  \BibitemOpen
  \bibfield  {author} {\bibinfo {author} {\bibfnamefont {A.~U.}\ \bibnamefont
  {Abeysekara}} \emph {et~al.} (\bibinfo {collaboration} {HAWC}),\ }\href
  {https://doi.org/10.3847/1538-4357/aa7555} {\bibfield  {journal} {\bibinfo
  {journal} {Astrophys. J.}\ }\textbf {\bibinfo {volume} {843}},\ \bibinfo
  {pages} {39} (\bibinfo {year} {2017}{\natexlab{a}})},\ \Eprint
  {https://arxiv.org/abs/1701.01778} {arXiv:1701.01778 [astro-ph.HE]}
  \BibitemShut {NoStop}%
\bibitem [{\citenamefont {Tavani}\ \emph {et~al.}(2018)\citenamefont {Tavani}
  \emph {et~al.}}]{DeAngelis:2017gra}%
  \BibitemOpen
  \bibfield  {author} {\bibinfo {author} {\bibfnamefont {M.}~\bibnamefont
  {Tavani}} \emph {et~al.} (\bibinfo {collaboration} {e-ASTROGAM}),\ }\href
  {https://doi.org/10.1016/j.jheap.2018.07.001} {\bibfield  {journal} {\bibinfo
   {journal} {JHEAp}\ }\textbf {\bibinfo {volume} {19}},\ \bibinfo {pages} {1}
  (\bibinfo {year} {2018})},\ \Eprint {https://arxiv.org/abs/1711.01265}
  {arXiv:1711.01265 [astro-ph.HE]} \BibitemShut {NoStop}%
\bibitem [{\citenamefont {Koglin}\ \emph {et~al.}(2009)\citenamefont {Koglin},
  \citenamefont {An}, \citenamefont {L.~Blaedel}, \citenamefont {F.~Brejnholt},
  \citenamefont {E.~Christensen}, \citenamefont {W.~Craig}, \citenamefont
  {Decker}, \citenamefont {J.~Hailey}, \citenamefont {Hale}, \citenamefont
  {A~Harrison}, \citenamefont {P.~Jensen}, \citenamefont {Madsen},
  \citenamefont {Mori}, \citenamefont {Pivovaroff}, \citenamefont {Tajiri},\
  and\ \citenamefont {W.~Zhang}}]{NuSTAR}%
  \BibitemOpen
  \bibfield  {author} {\bibinfo {author} {\bibfnamefont {J.}~\bibnamefont
  {Koglin}}, \bibinfo {author} {\bibfnamefont {H.}~\bibnamefont {An}}, \bibinfo
  {author} {\bibfnamefont {K.}~\bibnamefont {L.~Blaedel}}, \bibinfo {author}
  {\bibfnamefont {N.}~\bibnamefont {F.~Brejnholt}}, \bibinfo {author}
  {\bibfnamefont {F.}~\bibnamefont {E.~Christensen}}, \bibinfo {author}
  {\bibfnamefont {W.}~\bibnamefont {W.~Craig}}, \bibinfo {author}
  {\bibfnamefont {T.}~\bibnamefont {Decker}}, \bibinfo {author} {\bibfnamefont
  {C.}~\bibnamefont {J.~Hailey}}, \bibinfo {author} {\bibfnamefont
  {L.}~\bibnamefont {Hale}}, \bibinfo {author} {\bibfnamefont {F.}~\bibnamefont
  {A~Harrison}}, \bibinfo {author} {\bibfnamefont {C.}~\bibnamefont
  {P.~Jensen}}, \bibinfo {author} {\bibfnamefont {K.}~\bibnamefont {Madsen}},
  \bibinfo {author} {\bibfnamefont {K.}~\bibnamefont {Mori}}, \bibinfo {author}
  {\bibfnamefont {M.}~\bibnamefont {Pivovaroff}}, \bibinfo {author}
  {\bibfnamefont {G.}~\bibnamefont {Tajiri}},\ and\ \bibinfo {author}
  {\bibfnamefont {W.}~\bibnamefont {W.~Zhang}}\ }(\bibinfo {year}
  {2009})\BibitemShut {NoStop}%
\bibitem [{Swi()}]{SwiftSensitivity}%
  \BibitemOpen
  \href@noop {} {\bibinfo {title} {{Swift's X-Ray Telescope (XRT)}}},\ \bibinfo
  {howpublished}
  {\url{https://swift.gsfc.nasa.gov/about_swift/xrt_desc.html}},\ \bibinfo
  {note} {accessed: 2019-03-07}\BibitemShut {NoStop}%
\bibitem [{\citenamefont {{Hassan}}\ \emph {et~al.}(2017)\citenamefont
  {{Hassan}}, \citenamefont {{Arrabito}}, \citenamefont {{Bernl{\"o}hr}},
  \citenamefont {{Bregeon}}, \citenamefont {{Cortina}}, \citenamefont
  {{Cumani}}, \citenamefont {{Di Pierro}}, \citenamefont {{Falceta-Goncalves}},
  \citenamefont {{Lang}}, \citenamefont {{Hinton}}, \citenamefont {{Jogler}},
  \citenamefont {{Maier}}, \citenamefont {{Moralejo}}, \citenamefont
  {{Morselli}}, \citenamefont {{Todero Peixoto}},\ and\ \citenamefont
  {{Wood}}}]{2017APh....93...76H}%
  \BibitemOpen
  \bibfield  {author} {\bibinfo {author} {\bibfnamefont {T.}~\bibnamefont
  {{Hassan}}}, \bibinfo {author} {\bibfnamefont {L.}~\bibnamefont
  {{Arrabito}}}, \bibinfo {author} {\bibfnamefont {K.}~\bibnamefont
  {{Bernl{\"o}hr}}}, \bibinfo {author} {\bibfnamefont {J.}~\bibnamefont
  {{Bregeon}}}, \bibinfo {author} {\bibfnamefont {J.}~\bibnamefont
  {{Cortina}}}, \bibinfo {author} {\bibfnamefont {P.}~\bibnamefont {{Cumani}}},
  \bibinfo {author} {\bibfnamefont {F.}~\bibnamefont {{Di Pierro}}}, \bibinfo
  {author} {\bibfnamefont {D.}~\bibnamefont {{Falceta-Goncalves}}}, \bibinfo
  {author} {\bibfnamefont {R.~G.}\ \bibnamefont {{Lang}}}, \bibinfo {author}
  {\bibfnamefont {J.}~\bibnamefont {{Hinton}}}, \bibinfo {author}
  {\bibfnamefont {T.}~\bibnamefont {{Jogler}}}, \bibinfo {author}
  {\bibfnamefont {G.}~\bibnamefont {{Maier}}}, \bibinfo {author} {\bibfnamefont
  {A.}~\bibnamefont {{Moralejo}}}, \bibinfo {author} {\bibfnamefont
  {A.}~\bibnamefont {{Morselli}}}, \bibinfo {author} {\bibfnamefont {C.~J.}\
  \bibnamefont {{Todero Peixoto}}},\ and\ \bibinfo {author} {\bibfnamefont
  {M.}~\bibnamefont {{Wood}}},\ }\href
  {https://doi.org/10.1016/j.astropartphys.2017.05.001} {\bibfield  {journal}
  {\bibinfo  {journal} {Astroparticle Physics}\ }\textbf {\bibinfo {volume}
  {93}},\ \bibinfo {pages} {76} (\bibinfo {year} {2017})},\ \Eprint
  {https://arxiv.org/abs/1705.01790} {arXiv:1705.01790 [astro-ph.IM]}
  \BibitemShut {NoStop}%
\bibitem [{\citenamefont {Albert}\ \emph {et~al.}(2019)\citenamefont {Albert}
  \emph {et~al.}}]{Albert:2019afb}%
  \BibitemOpen
  \bibfield  {author} {\bibinfo {author} {\bibfnamefont {A.}~\bibnamefont
  {Albert}} \emph {et~al.} (\bibinfo {collaboration} {SGSO}),\ }\href@noop {}
  {\  (\bibinfo {year} {2019})},\ \Eprint {https://arxiv.org/abs/1902.08429}
  {arXiv:1902.08429 [astro-ph.HE]} \BibitemShut {NoStop}%
\bibitem [{\citenamefont {Caputo}\ \emph {et~al.}(2018)\citenamefont {Caputo},
  \citenamefont {Kislat},\ and\ \citenamefont {Racusin}}]{Caputo:2017tqk}%
  \BibitemOpen
  \bibfield  {author} {\bibinfo {author} {\bibfnamefont {R.}~\bibnamefont
  {Caputo}}, \bibinfo {author} {\bibfnamefont {F.}~\bibnamefont {Kislat}},\
  and\ \bibinfo {author} {\bibfnamefont {J.}~\bibnamefont {Racusin}} (\bibinfo
  {collaboration} {{AMEGO Team}}),\ }\bibfield  {booktitle} {\emph {\bibinfo
  {booktitle} {{Contributions to the 35th International Cosmic Ray Conference
  (ICRC 2017)}}},\ }\href {https://doi.org/10.22323/1.301.0783} {\bibfield
  {journal} {\bibinfo  {journal} {PoS}\ }\textbf {\bibinfo {volume}
  {ICRC2017}},\ \bibinfo {pages} {783} (\bibinfo {year} {2018})}\BibitemShut
  {NoStop}%
\bibitem [{TAP()}]{TAPSensitivity}%
  \BibitemOpen
  \href@noop {} {\bibinfo {title} {{TAP Instruments}}},\ \bibinfo
  {howpublished} {\url{https://asd.gsfc.nasa.gov/tap/instruments.html}},\
  \bibinfo {note} {accessed: 2019-03-07}\BibitemShut {NoStop}%
\bibitem [{\citenamefont {Ray}\ \emph {et~al.}(2019)\citenamefont {Ray} \emph
  {et~al.}}]{Ray:2019pxr}%
  \BibitemOpen
  \bibfield  {author} {\bibinfo {author} {\bibfnamefont {P.~S.}\ \bibnamefont
  {Ray}} \emph {et~al.} (\bibinfo {collaboration} {STROBE-X Science Working
  Group}),\ }\href@noop {} {\  (\bibinfo {year} {2019})},\ \Eprint
  {https://arxiv.org/abs/1903.03035} {arXiv:1903.03035 [astro-ph.IM]}
  \BibitemShut {NoStop}%
\bibitem [{\citenamefont {{Racusin}}(2016)}]{2016LPICo1962.4081R}%
  \BibitemOpen
  \bibfield  {author} {\bibinfo {author} {\bibfnamefont {J.~L.}\ \bibnamefont
  {{Racusin}}} (\bibinfo {collaboration} {TAO Team}),\ }in\ \href@noop {}
  {\emph {\bibinfo {booktitle} {Eighth Huntsville Gamma-Ray Burst
  Symposium}}},\ \bibinfo {series} {LPI Contributions}, Vol.\ \bibinfo {volume}
  {1962}\ (\bibinfo {year} {2016})\ p.\ \bibinfo {pages} {4081}\BibitemShut
  {NoStop}%
\bibitem [{\citenamefont {Aartsen}\ \emph
  {et~al.}(2018{\natexlab{b}})\citenamefont {Aartsen} \emph
  {et~al.}}]{IceCube:2018cha}%
  \BibitemOpen
  \bibfield  {author} {\bibinfo {author} {\bibfnamefont {M.~G.}\ \bibnamefont
  {Aartsen}} \emph {et~al.} (\bibinfo {collaboration} {IceCube}),\ }\href
  {https://doi.org/10.1126/science.aat2890} {\bibfield  {journal} {\bibinfo
  {journal} {Science}\ }\textbf {\bibinfo {volume} {361}},\ \bibinfo {pages}
  {147} (\bibinfo {year} {2018}{\natexlab{b}})},\ \Eprint
  {https://arxiv.org/abs/1807.08794} {arXiv:1807.08794 [astro-ph.HE]}
  \BibitemShut {NoStop}%
\bibitem [{\citenamefont {Aartsen}\ \emph
  {et~al.}(2017{\natexlab{a}})\citenamefont {Aartsen} \emph
  {et~al.}}]{Aartsen:2016oji}%
  \BibitemOpen
  \bibfield  {author} {\bibinfo {author} {\bibfnamefont {M.~G.}\ \bibnamefont
  {Aartsen}} \emph {et~al.} (\bibinfo {collaboration} {IceCube}),\ }\href
  {https://doi.org/10.3847/1538-4357/835/2/151} {\bibfield  {journal} {\bibinfo
   {journal} {Astrophys. J.}\ }\textbf {\bibinfo {volume} {835}},\ \bibinfo
  {pages} {151} (\bibinfo {year} {2017}{\natexlab{a}})},\ \Eprint
  {https://arxiv.org/abs/1609.04981} {arXiv:1609.04981 [astro-ph.HE]}
  \BibitemShut {NoStop}%
\bibitem [{\citenamefont {Aartsen}\ \emph
  {et~al.}(2013{\natexlab{a}})\citenamefont {Aartsen} \emph
  {et~al.}}]{Aartsen:2013bka}%
  \BibitemOpen
  \bibfield  {author} {\bibinfo {author} {\bibfnamefont {M.~G.}\ \bibnamefont
  {Aartsen}} \emph {et~al.} (\bibinfo {collaboration} {IceCube}),\ }\href
  {https://doi.org/10.1103/PhysRevLett.111.021103} {\bibfield  {journal}
  {\bibinfo  {journal} {Phys. Rev. Lett.}\ }\textbf {\bibinfo {volume} {111}},\
  \bibinfo {pages} {021103} (\bibinfo {year} {2013}{\natexlab{a}})},\ \Eprint
  {https://arxiv.org/abs/1304.5356} {arXiv:1304.5356 [astro-ph.HE]}
  \BibitemShut {NoStop}%
\bibitem [{\citenamefont {Aartsen}\ \emph
  {et~al.}(2013{\natexlab{b}})\citenamefont {Aartsen} \emph
  {et~al.}}]{Aartsen:2013jdh}%
  \BibitemOpen
  \bibfield  {author} {\bibinfo {author} {\bibfnamefont {M.~G.}\ \bibnamefont
  {Aartsen}} \emph {et~al.} (\bibinfo {collaboration} {IceCube}),\ }\href
  {https://doi.org/10.1126/science.1242856} {\bibfield  {journal} {\bibinfo
  {journal} {Science}\ }\textbf {\bibinfo {volume} {342}},\ \bibinfo {pages}
  {1242856} (\bibinfo {year} {2013}{\natexlab{b}})},\ \Eprint
  {https://arxiv.org/abs/1311.5238} {arXiv:1311.5238 [astro-ph.HE]}
  \BibitemShut {NoStop}%
\bibitem [{\citenamefont {Aartsen}\ \emph {et~al.}(2015)\citenamefont {Aartsen}
  \emph {et~al.}}]{Aartsen:2015wto}%
  \BibitemOpen
  \bibfield  {author} {\bibinfo {author} {\bibfnamefont {M.~G.}\ \bibnamefont
  {Aartsen}} \emph {et~al.} (\bibinfo {collaboration} {IceCube}),\ }\href
  {https://doi.org/10.1088/0004-637X/807/1/46} {\bibfield  {journal} {\bibinfo
  {journal} {Astrophys. J.}\ }\textbf {\bibinfo {volume} {807}},\ \bibinfo
  {pages} {46} (\bibinfo {year} {2015})},\ \Eprint
  {https://arxiv.org/abs/1503.00598} {arXiv:1503.00598 [astro-ph.HE]}
  \BibitemShut {NoStop}%
\bibitem [{\citenamefont {Aartsen}\ \emph {et~al.}(2019)\citenamefont {Aartsen}
  \emph {et~al.}}]{Aartsen:2018fpd}%
  \BibitemOpen
  \bibfield  {author} {\bibinfo {author} {\bibfnamefont {M.~G.}\ \bibnamefont
  {Aartsen}} \emph {et~al.} (\bibinfo {collaboration} {IceCube}),\ }\href
  {https://doi.org/10.1103/PhysRevLett.122.051102} {\bibfield  {journal}
  {\bibinfo  {journal} {Phys. Rev. Lett.}\ }\textbf {\bibinfo {volume} {122}},\
  \bibinfo {pages} {051102} (\bibinfo {year} {2019})},\ \Eprint
  {https://arxiv.org/abs/1807.11492} {arXiv:1807.11492 [astro-ph.HE]}
  \BibitemShut {NoStop}%
\bibitem [{\citenamefont {Albert}\ \emph
  {et~al.}(2018{\natexlab{a}})\citenamefont {Albert} \emph
  {et~al.}}]{Albert:2018vxw}%
  \BibitemOpen
  \bibfield  {author} {\bibinfo {author} {\bibfnamefont {A.}~\bibnamefont
  {Albert}} \emph {et~al.} (\bibinfo {collaboration} {ANTARES, IceCube}),\
  }\href {https://doi.org/10.3847/2041-8213/aaeecf} {\bibfield  {journal}
  {\bibinfo  {journal} {Astrophys. J.}\ }\textbf {\bibinfo {volume} {868}},\
  \bibinfo {pages} {L20} (\bibinfo {year} {2018}{\natexlab{a}})},\ \Eprint
  {https://arxiv.org/abs/1808.03531} {arXiv:1808.03531 [astro-ph.HE]}
  \BibitemShut {NoStop}%
\bibitem [{\citenamefont {{Waxman}}\ and\ \citenamefont
  {{Bahcall}}(1999)}]{1999PhRvD..59b3002W}%
  \BibitemOpen
  \bibfield  {author} {\bibinfo {author} {\bibfnamefont {E.}~\bibnamefont
  {{Waxman}}}\ and\ \bibinfo {author} {\bibfnamefont {J.}~\bibnamefont
  {{Bahcall}}},\ }\href {https://doi.org/10.1103/PhysRevD.59.023002} {\bibfield
   {journal} {\bibinfo  {journal} {\prd}\ }\textbf {\bibinfo {volume} {59}},\
  \bibinfo {eid} {023002} (\bibinfo {year} {1999})},\ \Eprint
  {https://arxiv.org/abs/hep-ph/9807282} {arXiv:hep-ph/9807282 [hep-ph]}
  \BibitemShut {NoStop}%
\bibitem [{\citenamefont {Murase}\ \emph {et~al.}(2013)\citenamefont {Murase},
  \citenamefont {Ahlers},\ and\ \citenamefont {Lacki}}]{Murase:2013rfa}%
  \BibitemOpen
  \bibfield  {author} {\bibinfo {author} {\bibfnamefont {K.}~\bibnamefont
  {Murase}}, \bibinfo {author} {\bibfnamefont {M.}~\bibnamefont {Ahlers}},\
  and\ \bibinfo {author} {\bibfnamefont {B.~C.}\ \bibnamefont {Lacki}},\ }\href
  {https://doi.org/10.1103/PhysRevD.88.121301} {\bibfield  {journal} {\bibinfo
  {journal} {Phys. Rev.}\ }\textbf {\bibinfo {volume} {D88}},\ \bibinfo {pages}
  {121301} (\bibinfo {year} {2013})},\ \Eprint
  {https://arxiv.org/abs/1306.3417} {arXiv:1306.3417 [astro-ph.HE]}
  \BibitemShut {NoStop}%
\bibitem [{\citenamefont {Murase}\ and\ \citenamefont
  {Waxman}(2016)}]{Murase:2016gly}%
  \BibitemOpen
  \bibfield  {author} {\bibinfo {author} {\bibfnamefont {K.}~\bibnamefont
  {Murase}}\ and\ \bibinfo {author} {\bibfnamefont {E.}~\bibnamefont
  {Waxman}},\ }\href {https://doi.org/10.1103/PhysRevD.94.103006} {\bibfield
  {journal} {\bibinfo  {journal} {Phys. Rev.}\ }\textbf {\bibinfo {volume}
  {D94}},\ \bibinfo {pages} {103006} (\bibinfo {year} {2016})},\ \Eprint
  {https://arxiv.org/abs/1607.01601} {arXiv:1607.01601 [astro-ph.HE]}
  \BibitemShut {NoStop}%
\bibitem [{\citenamefont {Murase}\ \emph {et~al.}(2016)\citenamefont {Murase},
  \citenamefont {Guetta},\ and\ \citenamefont {Ahlers}}]{Murase:2015xka}%
  \BibitemOpen
  \bibfield  {author} {\bibinfo {author} {\bibfnamefont {K.}~\bibnamefont
  {Murase}}, \bibinfo {author} {\bibfnamefont {D.}~\bibnamefont {Guetta}},\
  and\ \bibinfo {author} {\bibfnamefont {M.}~\bibnamefont {Ahlers}},\ }\href
  {https://doi.org/10.1103/PhysRevLett.116.071101} {\bibfield  {journal}
  {\bibinfo  {journal} {Phys. Rev. Lett.}\ }\textbf {\bibinfo {volume} {116}},\
  \bibinfo {pages} {071101} (\bibinfo {year} {2016})},\ \Eprint
  {https://arxiv.org/abs/1509.00805} {arXiv:1509.00805 [astro-ph.HE]}
  \BibitemShut {NoStop}%
\bibitem [{\citenamefont {Bechtol}\ \emph {et~al.}(2017)\citenamefont
  {Bechtol}, \citenamefont {Ahlers}, \citenamefont {Di~Mauro}, \citenamefont
  {Ajello},\ and\ \citenamefont {Vandenbroucke}}]{Bechtol:2015uqb}%
  \BibitemOpen
  \bibfield  {author} {\bibinfo {author} {\bibfnamefont {K.}~\bibnamefont
  {Bechtol}}, \bibinfo {author} {\bibfnamefont {M.}~\bibnamefont {Ahlers}},
  \bibinfo {author} {\bibfnamefont {M.}~\bibnamefont {Di~Mauro}}, \bibinfo
  {author} {\bibfnamefont {M.}~\bibnamefont {Ajello}},\ and\ \bibinfo {author}
  {\bibfnamefont {J.}~\bibnamefont {Vandenbroucke}},\ }\href
  {https://doi.org/10.3847/1538-4357/836/1/47} {\bibfield  {journal} {\bibinfo
  {journal} {Astrophys. J.}\ }\textbf {\bibinfo {volume} {836}},\ \bibinfo
  {pages} {47} (\bibinfo {year} {2017})},\ \Eprint
  {https://arxiv.org/abs/1511.00688} {arXiv:1511.00688 [astro-ph.HE]}
  \BibitemShut {NoStop}%
\bibitem [{\citenamefont {Xiao}\ \emph {et~al.}(2016)\citenamefont {Xiao},
  \citenamefont {Mészáros}, \citenamefont {Murase},\ and\ \citenamefont
  {Dai}}]{Xiao:2016rvd}%
  \BibitemOpen
  \bibfield  {author} {\bibinfo {author} {\bibfnamefont {D.}~\bibnamefont
  {Xiao}}, \bibinfo {author} {\bibfnamefont {P.}~\bibnamefont {Mészáros}},
  \bibinfo {author} {\bibfnamefont {K.}~\bibnamefont {Murase}},\ and\ \bibinfo
  {author} {\bibfnamefont {Z.-g.}\ \bibnamefont {Dai}},\ }\href
  {https://doi.org/10.3847/0004-637X/826/2/133} {\bibfield  {journal} {\bibinfo
   {journal} {Astrophys. J.}\ }\textbf {\bibinfo {volume} {826}},\ \bibinfo
  {pages} {133} (\bibinfo {year} {2016})},\ \Eprint
  {https://arxiv.org/abs/1604.08131} {arXiv:1604.08131 [astro-ph.HE]}
  \BibitemShut {NoStop}%
\bibitem [{\citenamefont {Padovani}\ \emph {et~al.}(2015)\citenamefont
  {Padovani}, \citenamefont {Petropoulou}, \citenamefont {Giommi},\ and\
  \citenamefont {Resconi}}]{Padovani:2015mba}%
  \BibitemOpen
  \bibfield  {author} {\bibinfo {author} {\bibfnamefont {P.}~\bibnamefont
  {Padovani}}, \bibinfo {author} {\bibfnamefont {M.}~\bibnamefont
  {Petropoulou}}, \bibinfo {author} {\bibfnamefont {P.}~\bibnamefont
  {Giommi}},\ and\ \bibinfo {author} {\bibfnamefont {E.}~\bibnamefont
  {Resconi}},\ }\href {https://doi.org/10.1093/mnras/stv1467} {\bibfield
  {journal} {\bibinfo  {journal} {Mon. Not. Roy. Astron. Soc.}\ }\textbf
  {\bibinfo {volume} {452}},\ \bibinfo {pages} {1877} (\bibinfo {year}
  {2015})},\ \Eprint {https://arxiv.org/abs/1506.09135} {arXiv:1506.09135
  [astro-ph.HE]} \BibitemShut {NoStop}%
\bibitem [{\citenamefont {Palladino}\ \emph {et~al.}(2018)\citenamefont
  {Palladino}, \citenamefont {Fedynitch}, \citenamefont {Rasmussen},\ and\
  \citenamefont {Taylor}}]{Palladino:2018bqf}%
  \BibitemOpen
  \bibfield  {author} {\bibinfo {author} {\bibfnamefont {A.}~\bibnamefont
  {Palladino}}, \bibinfo {author} {\bibfnamefont {A.}~\bibnamefont
  {Fedynitch}}, \bibinfo {author} {\bibfnamefont {R.~W.}\ \bibnamefont
  {Rasmussen}},\ and\ \bibinfo {author} {\bibfnamefont {A.~M.}\ \bibnamefont
  {Taylor}},\ }\href@noop {} {\  (\bibinfo {year} {2018})},\ \Eprint
  {https://arxiv.org/abs/1812.04685} {arXiv:1812.04685 [astro-ph.HE]}
  \BibitemShut {NoStop}%
\bibitem [{\citenamefont {{Murase}}(2017)}]{Murase2017}%
  \BibitemOpen
  \bibfield  {author} {\bibinfo {author} {\bibfnamefont {K.}~\bibnamefont
  {{Murase}}},\ }\bibinfo {title} {{Active Galactic Nuclei as High-Energy
  Neutrino Sources}},\ in\ \href {https://doi.org/10.1142/9789814759410_0002}
  {\emph {\bibinfo {booktitle} {Neutrino Astronomy: Current Status, Future
  Prospects. Edited by Thomas Gaisser Albrecht Karle. Published by World
  Scientific Publishing Co. Pte. Ltd., 2017. ISBN \#9789814759410, pp.
  15-31}}}\ (\bibinfo {year} {2017})\ pp.\ \bibinfo {pages}
  {15--31}\BibitemShut {NoStop}%
\bibitem [{\citenamefont {Aartsen}\ \emph
  {et~al.}(2017{\natexlab{b}})\citenamefont {Aartsen} \emph
  {et~al.}}]{Aartsen:2016lir}%
  \BibitemOpen
  \bibfield  {author} {\bibinfo {author} {\bibfnamefont {M.~G.}\ \bibnamefont
  {Aartsen}} \emph {et~al.} (\bibinfo {collaboration} {IceCube}),\ }\href
  {https://doi.org/10.3847/1538-4357/835/1/45} {\bibfield  {journal} {\bibinfo
  {journal} {Astrophys. J.}\ }\textbf {\bibinfo {volume} {835}},\ \bibinfo
  {pages} {45} (\bibinfo {year} {2017}{\natexlab{b}})},\ \Eprint
  {https://arxiv.org/abs/1611.03874} {arXiv:1611.03874 [astro-ph.HE]}
  \BibitemShut {NoStop}%
\bibitem [{\citenamefont {{Aartsen}}\ \emph {et~al.}(2017)\citenamefont
  {{Aartsen}} \emph {et~al.}}]{2017arXiv171001179I}%
  \BibitemOpen
  \bibfield  {author} {\bibinfo {author} {\bibfnamefont {M.~G.}\ \bibnamefont
  {{Aartsen}}} \emph {et~al.} (\bibinfo {collaboration} {IceCube}),\
  }\href@noop {} {\bibfield  {journal} {\bibinfo  {journal} {arXiv e-prints}\
  ,\ \bibinfo {eid} {arXiv:1710.01179}} (\bibinfo {year} {2017})},\ \Eprint
  {https://arxiv.org/abs/1710.01179} {arXiv:1710.01179 [astro-ph.HE]}
  \BibitemShut {NoStop}%
\bibitem [{\citenamefont {{Padovani}}\ and\ \citenamefont
  {{Resconi}}(2014)}]{Padovani2014}%
  \BibitemOpen
  \bibfield  {author} {\bibinfo {author} {\bibfnamefont {P.}~\bibnamefont
  {{Padovani}}}\ and\ \bibinfo {author} {\bibfnamefont {E.}~\bibnamefont
  {{Resconi}}},\ }\href {https://doi.org/10.1093/mnras/stu1166} {\bibfield
  {journal} {\bibinfo  {journal} {\mnras}\ }\textbf {\bibinfo {volume} {443}},\
  \bibinfo {pages} {474} (\bibinfo {year} {2014})},\ \Eprint
  {https://arxiv.org/abs/1406.0376} {arXiv:1406.0376 [astro-ph.HE]}
  \BibitemShut {NoStop}%
\bibitem [{\citenamefont {{Padovani}}\ \emph {et~al.}(2016)\citenamefont
  {{Padovani}}, \citenamefont {{Resconi}}, \citenamefont {{Giommi}},
  \citenamefont {{Arsioli}},\ and\ \citenamefont {{Chang}}}]{Padovani2016}%
  \BibitemOpen
  \bibfield  {author} {\bibinfo {author} {\bibfnamefont {P.}~\bibnamefont
  {{Padovani}}}, \bibinfo {author} {\bibfnamefont {E.}~\bibnamefont
  {{Resconi}}}, \bibinfo {author} {\bibfnamefont {P.}~\bibnamefont {{Giommi}}},
  \bibinfo {author} {\bibfnamefont {B.}~\bibnamefont {{Arsioli}}},\ and\
  \bibinfo {author} {\bibfnamefont {Y.~L.}\ \bibnamefont {{Chang}}},\ }\href
  {https://doi.org/10.1093/mnras/stw228} {\bibfield  {journal} {\bibinfo
  {journal} {\mnras}\ }\textbf {\bibinfo {volume} {457}},\ \bibinfo {pages}
  {3582} (\bibinfo {year} {2016})},\ \Eprint {https://arxiv.org/abs/1601.06550}
  {arXiv:1601.06550 [astro-ph.HE]} \BibitemShut {NoStop}%
\bibitem [{\citenamefont {Kadler}\ \emph {et~al.}(2016)\citenamefont {Kadler}
  \emph {et~al.}}]{Kadler:2016ygj}%
  \BibitemOpen
  \bibfield  {author} {\bibinfo {author} {\bibfnamefont {M.}~\bibnamefont
  {Kadler}} \emph {et~al.},\ }\href {https://doi.org/10.1038/nphys3715,
  10.1038/NPHYS3715} {\bibfield  {journal} {\bibinfo  {journal} {Nature Phys.}\
  }\textbf {\bibinfo {volume} {12}},\ \bibinfo {pages} {807} (\bibinfo {year}
  {2016})},\ \Eprint {https://arxiv.org/abs/1602.02012} {arXiv:1602.02012
  [astro-ph.HE]} \BibitemShut {NoStop}%
\bibitem [{\citenamefont {{Gu{\'e}pin}}\ and\ \citenamefont
  {{Kotera}}(2017)}]{Guepin2017}%
  \BibitemOpen
  \bibfield  {author} {\bibinfo {author} {\bibfnamefont {C.}~\bibnamefont
  {{Gu{\'e}pin}}}\ and\ \bibinfo {author} {\bibfnamefont {K.}~\bibnamefont
  {{Kotera}}},\ }\href {https://doi.org/10.1051/0004-6361/201630326} {\bibfield
   {journal} {\bibinfo  {journal} {\aap}\ }\textbf {\bibinfo {volume} {603}},\
  \bibinfo {eid} {A76} (\bibinfo {year} {2017})},\ \Eprint
  {https://arxiv.org/abs/1701.07038} {arXiv:1701.07038 [astro-ph.HE]}
  \BibitemShut {NoStop}%
\bibitem [{\citenamefont {Albert}\ \emph
  {et~al.}(2018{\natexlab{b}})\citenamefont {Albert} \emph
  {et~al.}}]{Albert:2018kjg}%
  \BibitemOpen
  \bibfield  {author} {\bibinfo {author} {\bibfnamefont {A.}~\bibnamefont
  {Albert}} \emph {et~al.} (\bibinfo {collaboration} {ANTARES}),\ }\href
  {https://doi.org/10.3847/2041-8213/aad8c0} {\bibfield  {journal} {\bibinfo
  {journal} {Astrophys. J.}\ }\textbf {\bibinfo {volume} {863}},\ \bibinfo
  {pages} {L30} (\bibinfo {year} {2018}{\natexlab{b}})},\ \Eprint
  {https://arxiv.org/abs/1807.04309} {arXiv:1807.04309 [astro-ph.HE]}
  \BibitemShut {NoStop}%
\bibitem [{\citenamefont {Ansoldi}\ \emph {et~al.}(2018)\citenamefont {Ansoldi}
  \emph {et~al.}}]{Ahnen:2018mvi}%
  \BibitemOpen
  \bibfield  {author} {\bibinfo {author} {\bibfnamefont {S.}~\bibnamefont
  {Ansoldi}} \emph {et~al.} (\bibinfo {collaboration} {MAGIC}),\ }\bibfield
  {journal} {\bibinfo  {journal} {Astrophys. J. Lett.}\ }\href
  {https://doi.org/10.3847/2041-8213/aad083} {10.3847/2041-8213/aad083}
  (\bibinfo {year} {2018}),\ \Eprint {https://arxiv.org/abs/1807.04300}
  {arXiv:1807.04300 [astro-ph.HE]} \BibitemShut {NoStop}%
\bibitem [{\citenamefont {Abeysekara}\ \emph {et~al.}(2018)\citenamefont
  {Abeysekara} \emph {et~al.}}]{Abeysekara:2018oub}%
  \BibitemOpen
  \bibfield  {author} {\bibinfo {author} {\bibfnamefont {A.~U.}\ \bibnamefont
  {Abeysekara}} \emph {et~al.} (\bibinfo {collaboration} {VERITAS}),\ }\href
  {https://doi.org/10.3847/2041-8213/aad053} {\bibfield  {journal} {\bibinfo
  {journal} {Astrophys. J.}\ }\textbf {\bibinfo {volume} {861}},\ \bibinfo
  {pages} {L20} (\bibinfo {year} {2018})},\ \Eprint
  {https://arxiv.org/abs/1807.04607} {arXiv:1807.04607 [astro-ph.HE]}
  \BibitemShut {NoStop}%
\bibitem [{\citenamefont {Keivani}\ \emph
  {et~al.}(2018{\natexlab{a}})\citenamefont {Keivani} \emph
  {et~al.}}]{Keivani:2018rnh}%
  \BibitemOpen
  \bibfield  {author} {\bibinfo {author} {\bibfnamefont {A.}~\bibnamefont
  {Keivani}} \emph {et~al.},\ }\href {https://doi.org/10.3847/1538-4357/aad59a}
  {\bibfield  {journal} {\bibinfo  {journal} {Astrophys. J.}\ }\textbf
  {\bibinfo {volume} {864}},\ \bibinfo {pages} {84} (\bibinfo {year}
  {2018}{\natexlab{a}})},\ \Eprint {https://arxiv.org/abs/1807.04537}
  {arXiv:1807.04537 [astro-ph.HE]} \BibitemShut {NoStop}%
\bibitem [{\citenamefont {Cerruti}\ \emph {et~al.}(2019)\citenamefont
  {Cerruti}, \citenamefont {Zech}, \citenamefont {Boisson}, \citenamefont
  {Emery}, \citenamefont {Inoue},\ and\ \citenamefont
  {Lenain}}]{Cerruti:2018tmc}%
  \BibitemOpen
  \bibfield  {author} {\bibinfo {author} {\bibfnamefont {M.}~\bibnamefont
  {Cerruti}}, \bibinfo {author} {\bibfnamefont {A.}~\bibnamefont {Zech}},
  \bibinfo {author} {\bibfnamefont {C.}~\bibnamefont {Boisson}}, \bibinfo
  {author} {\bibfnamefont {G.}~\bibnamefont {Emery}}, \bibinfo {author}
  {\bibfnamefont {S.}~\bibnamefont {Inoue}},\ and\ \bibinfo {author}
  {\bibfnamefont {J.~P.}\ \bibnamefont {Lenain}},\ }\href
  {https://doi.org/10.1093/mnrasl/sly210} {\bibfield  {journal} {\bibinfo
  {journal} {Mon. Not. Roy. Astron. Soc.}\ }\textbf {\bibinfo {volume} {483}},\
  \bibinfo {pages} {L12} (\bibinfo {year} {2019})},\ \Eprint
  {https://arxiv.org/abs/1807.04335} {arXiv:1807.04335 [astro-ph.HE]}
  \BibitemShut {NoStop}%
\bibitem [{\citenamefont {Gao}\ \emph {et~al.}(2019)\citenamefont {Gao},
  \citenamefont {Fedynitch}, \citenamefont {Winter},\ and\ \citenamefont
  {Pohl}}]{Gao:2018mnu}%
  \BibitemOpen
  \bibfield  {author} {\bibinfo {author} {\bibfnamefont {S.}~\bibnamefont
  {Gao}}, \bibinfo {author} {\bibfnamefont {A.}~\bibnamefont {Fedynitch}},
  \bibinfo {author} {\bibfnamefont {W.}~\bibnamefont {Winter}},\ and\ \bibinfo
  {author} {\bibfnamefont {M.}~\bibnamefont {Pohl}},\ }\href
  {https://doi.org/10.1038/s41550-018-0610-1} {\bibfield  {journal} {\bibinfo
  {journal} {Nat. Astron.}\ }\textbf {\bibinfo {volume} {3}},\ \bibinfo {pages}
  {88} (\bibinfo {year} {2019})},\ \Eprint {https://arxiv.org/abs/1807.04275}
  {arXiv:1807.04275 [astro-ph.HE]} \BibitemShut {NoStop}%
\bibitem [{\citenamefont {Padovani}\ \emph {et~al.}(2018)\citenamefont
  {Padovani}, \citenamefont {Giommi}, \citenamefont {Resconi}, \citenamefont
  {Glauch}, \citenamefont {Arsioli}, \citenamefont {Sahakyan},\ and\
  \citenamefont {Huber}}]{Padovani:2018acg}%
  \BibitemOpen
  \bibfield  {author} {\bibinfo {author} {\bibfnamefont {P.}~\bibnamefont
  {Padovani}}, \bibinfo {author} {\bibfnamefont {P.}~\bibnamefont {Giommi}},
  \bibinfo {author} {\bibfnamefont {E.}~\bibnamefont {Resconi}}, \bibinfo
  {author} {\bibfnamefont {T.}~\bibnamefont {Glauch}}, \bibinfo {author}
  {\bibfnamefont {B.}~\bibnamefont {Arsioli}}, \bibinfo {author} {\bibfnamefont
  {N.}~\bibnamefont {Sahakyan}},\ and\ \bibinfo {author} {\bibfnamefont
  {M.}~\bibnamefont {Huber}},\ }\href {https://doi.org/10.1093/mnras/sty1852}
  {\bibfield  {journal} {\bibinfo  {journal} {Mon. Not. Roy. Astron. Soc.}\
  }\textbf {\bibinfo {volume} {480}},\ \bibinfo {pages} {192} (\bibinfo {year}
  {2018})},\ \Eprint {https://arxiv.org/abs/1807.04461} {arXiv:1807.04461
  [astro-ph.HE]} \BibitemShut {NoStop}%
\bibitem [{\citenamefont {Garrappa}\ \emph {et~al.}(2019)\citenamefont
  {Garrappa} \emph {et~al.}}]{Aartsen:2019gxs}%
  \BibitemOpen
  \bibfield  {author} {\bibinfo {author} {\bibfnamefont {S.}~\bibnamefont
  {Garrappa}} \emph {et~al.} (\bibinfo {collaboration} {Fermi-LAT, ASAS-SN,
  IceCube}),\ }\href@noop {} {\  (\bibinfo {year} {2019})},\ \Eprint
  {https://arxiv.org/abs/1901.10806} {arXiv:1901.10806 [astro-ph.HE]}
  \BibitemShut {NoStop}%
\bibitem [{\citenamefont {Rodrigues}\ \emph {et~al.}(2018)\citenamefont
  {Rodrigues}, \citenamefont {Gao}, \citenamefont {Fedynitch}, \citenamefont
  {Palladino},\ and\ \citenamefont {Winter}}]{Rodrigues:2018tku}%
  \BibitemOpen
  \bibfield  {author} {\bibinfo {author} {\bibfnamefont {X.}~\bibnamefont
  {Rodrigues}}, \bibinfo {author} {\bibfnamefont {S.}~\bibnamefont {Gao}},
  \bibinfo {author} {\bibfnamefont {A.}~\bibnamefont {Fedynitch}}, \bibinfo
  {author} {\bibfnamefont {A.}~\bibnamefont {Palladino}},\ and\ \bibinfo
  {author} {\bibfnamefont {W.}~\bibnamefont {Winter}},\ }\href@noop {} {\
  (\bibinfo {year} {2018})},\ \Eprint {https://arxiv.org/abs/1812.05939}
  {arXiv:1812.05939 [astro-ph.HE]} \BibitemShut {NoStop}%
\bibitem [{\citenamefont {Reimer}\ \emph {et~al.}(2018)\citenamefont {Reimer},
  \citenamefont {Boettcher},\ and\ \citenamefont {Buson}}]{Reimer:2018vvw}%
  \BibitemOpen
  \bibfield  {author} {\bibinfo {author} {\bibfnamefont {A.}~\bibnamefont
  {Reimer}}, \bibinfo {author} {\bibfnamefont {M.}~\bibnamefont {Boettcher}},\
  and\ \bibinfo {author} {\bibfnamefont {S.}~\bibnamefont {Buson}},\
  }\href@noop {} {\  (\bibinfo {year} {2018})},\ \Eprint
  {https://arxiv.org/abs/1812.05654} {arXiv:1812.05654 [astro-ph.HE]}
  \BibitemShut {NoStop}%
\bibitem [{\citenamefont {{Petropoulou}}\ and\ \citenamefont
  {{Mastichiadis}}(2015)}]{PetroMast2015}%
  \BibitemOpen
  \bibfield  {author} {\bibinfo {author} {\bibfnamefont {M.}~\bibnamefont
  {{Petropoulou}}}\ and\ \bibinfo {author} {\bibfnamefont {A.}~\bibnamefont
  {{Mastichiadis}}},\ }\href {https://doi.org/10.1093/mnras/stu2364} {\bibfield
   {journal} {\bibinfo  {journal} {\mnras}\ }\textbf {\bibinfo {volume}
  {447}},\ \bibinfo {pages} {36} (\bibinfo {year} {2015})},\ \Eprint
  {https://arxiv.org/abs/1411.1908} {arXiv:1411.1908 [astro-ph.HE]}
  \BibitemShut {NoStop}%
\bibitem [{\citenamefont {{Santander}}(2017)}]{Santander:2016bvv}%
  \BibitemOpen
  \bibfield  {author} {\bibinfo {author} {\bibfnamefont {M.}~\bibnamefont
  {{Santander}}},\ }\bibinfo {title} {{The Dawn of Multi-Messenger
  Astronomy}},\ in\ \href {https://doi.org/10.1142/9789814759410_0009} {\emph
  {\bibinfo {booktitle} {Neutrino Astronomy: Current Status, Future Prospects.
  Edited by Thomas Gaisser Albrecht Karle. Published by World Scientific
  Publishing Co. Pte. Ltd., 2017. ISBN \#9789814759410, pp. 125-140}}}\
  (\bibinfo {year} {2017})\ pp.\ \bibinfo {pages} {125--140}\BibitemShut
  {NoStop}%
\bibitem [{\citenamefont {Aartsen}\ \emph
  {et~al.}(2017{\natexlab{a}})\citenamefont {Aartsen} \emph
  {et~al.}}]{Aartsen:2016lmt}%
  \BibitemOpen
  \bibfield  {author} {\bibinfo {author} {\bibfnamefont {M.~G.}\ \bibnamefont
  {Aartsen}} \emph {et~al.} (\bibinfo {collaboration} {IceCube}),\ }\href
  {https://doi.org/10.1016/j.astropartphys.2017.05.002} {\bibfield  {journal}
  {\bibinfo  {journal} {Astropart. Phys.}\ }\textbf {\bibinfo {volume} {92}},\
  \bibinfo {pages} {30} (\bibinfo {year} {2017}{\natexlab{a}})},\ \Eprint
  {https://arxiv.org/abs/1612.06028} {arXiv:1612.06028 [astro-ph.HE]}
  \BibitemShut {NoStop}%
\bibitem [{\citenamefont {{Ageron}}\ \emph {et~al.}(2011)\citenamefont
  {{Ageron}} \emph {et~al.}}]{2011NIMPA.656...11A}%
  \BibitemOpen
  \bibfield  {author} {\bibinfo {author} {\bibfnamefont {M.}~\bibnamefont
  {{Ageron}}} \emph {et~al.} (\bibinfo {collaboration} {ANTARES}),\ }\href
  {https://doi.org/10.1016/j.nima.2011.06.103} {\bibfield  {journal} {\bibinfo
  {journal} {Nuclear Instruments and Methods in Physics Research A}\ }\textbf
  {\bibinfo {volume} {656}},\ \bibinfo {pages} {11} (\bibinfo {year} {2011})},\
  \Eprint {https://arxiv.org/abs/1104.1607} {arXiv:1104.1607 [astro-ph.IM]}
  \BibitemShut {NoStop}%
\bibitem [{\citenamefont {Aynutdinov}\ \emph {et~al.}(2009)\citenamefont
  {Aynutdinov} \emph {et~al.}}]{Aynutdinov:2009zzb}%
  \BibitemOpen
  \bibfield  {author} {\bibinfo {author} {\bibfnamefont {V.}~\bibnamefont
  {Aynutdinov}} \emph {et~al.},\ }\bibfield  {booktitle} {\emph {\bibinfo
  {booktitle} {{Very large volume neutrino telescope for the Mediterranean Sea.
  Proceedings, 3rd International VLVnuT Workshop, Toulon, France, April 22-24,
  2008}}},\ }\href {https://doi.org/10.1016/j.nima.2008.12.012} {\bibfield
  {journal} {\bibinfo  {journal} {Nucl. Instrum. Meth.}\ }\textbf {\bibinfo
  {volume} {A602}},\ \bibinfo {pages} {14} (\bibinfo {year} {2009})},\ \Eprint
  {https://arxiv.org/abs/0811.1109} {arXiv:0811.1109 [astro-ph]} \BibitemShut
  {NoStop}%
\bibitem [{\citenamefont {Adrian-Martinez}\ \emph {et~al.}(2016)\citenamefont
  {Adrian-Martinez} \emph {et~al.}}]{Adrian-Martinez:2016fdl}%
  \BibitemOpen
  \bibfield  {author} {\bibinfo {author} {\bibfnamefont {S.}~\bibnamefont
  {Adrian-Martinez}} \emph {et~al.} (\bibinfo {collaboration} {KM3NeT}),\
  }\href {https://doi.org/10.1088/0954-3899/43/8/084001} {\bibfield  {journal}
  {\bibinfo  {journal} {J. Phys.}\ }\textbf {\bibinfo {volume} {G43}},\
  \bibinfo {pages} {084001} (\bibinfo {year} {2016})},\ \Eprint
  {https://arxiv.org/abs/1601.07459} {arXiv:1601.07459 [astro-ph.IM]}
  \BibitemShut {NoStop}%
\bibitem [{\citenamefont {Avrorin}\ \emph {et~al.}(2018)\citenamefont {Avrorin}
  \emph {et~al.}}]{Avrorin:2018ijk}%
  \BibitemOpen
  \bibfield  {author} {\bibinfo {author} {\bibfnamefont {A.~D.}\ \bibnamefont
  {Avrorin}} \emph {et~al.} (\bibinfo {collaboration} {Baikal-GVD}),\
  }\bibfield  {booktitle} {\emph {\bibinfo {booktitle} {{Proceedings, 20th
  International Seminar on High Energy Physics (Quarks 2018): Valday, Russia,
  May 27 -June 2, 2018}}},\ }\href
  {https://doi.org/10.1051/epjconf/201819101006} {\bibfield  {journal}
  {\bibinfo  {journal} {EPJ Web Conf.}\ }\textbf {\bibinfo {volume} {191}},\
  \bibinfo {pages} {01006} (\bibinfo {year} {2018})},\ \Eprint
  {https://arxiv.org/abs/1808.10353} {arXiv:1808.10353 [astro-ph.IM]}
  \BibitemShut {NoStop}%
\bibitem [{\citenamefont {Aartsen}\ \emph {et~al.}(2014)\citenamefont {Aartsen}
  \emph {et~al.}}]{Aartsen:2014njl}%
  \BibitemOpen
  \bibfield  {author} {\bibinfo {author} {\bibfnamefont {M.~G.}\ \bibnamefont
  {Aartsen}} \emph {et~al.} (\bibinfo {collaboration} {IceCube}),\ }\href@noop
  {} {\  (\bibinfo {year} {2014})},\ \Eprint {https://arxiv.org/abs/1412.5106}
  {arXiv:1412.5106 [astro-ph.HE]} \BibitemShut {NoStop}%
\bibitem [{\citenamefont {{Aiello}}\ \emph {et~al.}(2018)\citenamefont
  {{Aiello}} \emph {et~al.}}]{2018arXiv181008499T}%
  \BibitemOpen
  \bibfield  {author} {\bibinfo {author} {\bibfnamefont {S.}~\bibnamefont
  {{Aiello}}} \emph {et~al.} (\bibinfo {collaboration} {KM3NeT}),\ }\href@noop
  {} {\bibfield  {journal} {\bibinfo  {journal} {arXiv e-prints}\ ,\ \bibinfo
  {eid} {arXiv:1810.08499}} (\bibinfo {year} {2018})},\ \Eprint
  {https://arxiv.org/abs/1810.08499} {arXiv:1810.08499 [astro-ph.HE]}
  \BibitemShut {NoStop}%
\bibitem [{\citenamefont {van Santen}(2018)}]{vanSanten:2017chb}%
  \BibitemOpen
  \bibfield  {author} {\bibinfo {author} {\bibfnamefont {J.}~\bibnamefont {van
  Santen}} (\bibinfo {collaboration} {IceCube Gen2}),\ }\bibfield  {booktitle}
  {\emph {\bibinfo {booktitle} {{Contributions to the 35th International Cosmic
  Ray Conference (ICRC 2017)}}},\ }\href {https://doi.org/10.22323/1.301.0991}
  {\bibfield  {journal} {\bibinfo  {journal} {PoS}\ }\textbf {\bibinfo {volume}
  {ICRC2017}},\ \bibinfo {pages} {991} (\bibinfo {year} {2018})}\BibitemShut
  {NoStop}%
\bibitem [{\citenamefont {{Allison}}\ \emph {et~al.}(2012)\citenamefont
  {{Allison}} \emph {et~al.}}]{2012APh....35..457A}%
  \BibitemOpen
  \bibfield  {author} {\bibinfo {author} {\bibfnamefont {P.}~\bibnamefont
  {{Allison}}} \emph {et~al.} (\bibinfo {collaboration} {ARA}),\ }\href
  {https://doi.org/10.1016/j.astropartphys.2011.11.010} {\bibfield  {journal}
  {\bibinfo  {journal} {Astroparticle Physics}\ }\textbf {\bibinfo {volume}
  {35}},\ \bibinfo {pages} {457} (\bibinfo {year} {2012})},\ \Eprint
  {https://arxiv.org/abs/1105.2854} {arXiv:1105.2854 [astro-ph.IM]}
  \BibitemShut {NoStop}%
\bibitem [{\citenamefont {{Barwick}}\ \emph {et~al.}(2017)\citenamefont
  {{Barwick}} \emph {et~al.}}]{2017APh....90...50B}%
  \BibitemOpen
  \bibfield  {author} {\bibinfo {author} {\bibfnamefont {S.~W.}\ \bibnamefont
  {{Barwick}}} \emph {et~al.} (\bibinfo {collaboration} {ARIANNA}),\ }\href
  {https://doi.org/10.1016/j.astropartphys.2017.02.003} {\bibfield  {journal}
  {\bibinfo  {journal} {Astroparticle Physics}\ }\textbf {\bibinfo {volume}
  {90}},\ \bibinfo {pages} {50} (\bibinfo {year} {2017})},\ \Eprint
  {https://arxiv.org/abs/1612.04473} {arXiv:1612.04473 [astro-ph.IM]}
  \BibitemShut {NoStop}%
\bibitem [{\citenamefont {{Alvarez-Muniz}}\ \emph {et~al.}(2018)\citenamefont
  {{Alvarez-Muniz}} \emph {et~al.}}]{2018arXiv181009994G}%
  \BibitemOpen
  \bibfield  {author} {\bibinfo {author} {\bibfnamefont {J.}~\bibnamefont
  {{Alvarez-Muniz}}} \emph {et~al.} (\bibinfo {collaboration} {GRAND}),\
  }\href@noop {} {\bibfield  {journal} {\bibinfo  {journal} {arXiv e-prints}\
  ,\ \bibinfo {eid} {arXiv:1810.09994}} (\bibinfo {year} {2018})},\ \Eprint
  {https://arxiv.org/abs/1810.09994} {arXiv:1810.09994 [astro-ph.HE]}
  \BibitemShut {NoStop}%
\bibitem [{\citenamefont {Ackermann}\ \emph {et~al.}(2019)\citenamefont
  {Ackermann} \emph {et~al.}}]{nuwp}%
  \BibitemOpen
  \bibfield  {author} {\bibinfo {author} {\bibfnamefont {M.}~\bibnamefont
  {Ackermann}} \emph {et~al.},\ }\href@noop {} {\bibfield  {journal} {\bibinfo
  {journal} {{Astro2020: Astrophysics Uniquely Enabled by Observations of
  High-Energy Cosmic Neutrinos}}\ } (\bibinfo {year} {2019})}\BibitemShut
  {NoStop}%
\bibitem [{\citenamefont {Richards}\ \emph {et~al.}(2011)\citenamefont
  {Richards} \emph {et~al.}}]{Richards:2010su}%
  \BibitemOpen
  \bibfield  {author} {\bibinfo {author} {\bibfnamefont {J.~L.}\ \bibnamefont
  {Richards}} \emph {et~al.} (\bibinfo {collaboration} {OVRO}),\ }\href
  {https://doi.org/10.1088/0067-0049/194/2/29} {\bibfield  {journal} {\bibinfo
  {journal} {Astrophys. J. Suppl.}\ }\textbf {\bibinfo {volume} {194}},\
  \bibinfo {pages} {29} (\bibinfo {year} {2011})},\ \Eprint
  {https://arxiv.org/abs/1011.3111} {arXiv:1011.3111 [astro-ph.CO]}
  \BibitemShut {NoStop}%
\bibitem [{\citenamefont {Lister}\ \emph {et~al.}(2009)\citenamefont {Lister}
  \emph {et~al.}}]{Lister:2008bw}%
  \BibitemOpen
  \bibfield  {author} {\bibinfo {author} {\bibfnamefont {M.~L.}\ \bibnamefont
  {Lister}} \emph {et~al.} (\bibinfo {collaboration} {MOJAVE}),\ }\href
  {https://doi.org/10.1088/0004-6256/137/3/3718} {\bibfield  {journal}
  {\bibinfo  {journal} {Astron. J.}\ }\textbf {\bibinfo {volume} {137}},\
  \bibinfo {pages} {3718} (\bibinfo {year} {2009})},\ \Eprint
  {https://arxiv.org/abs/0812.3947} {arXiv:0812.3947 [astro-ph]} \BibitemShut
  {NoStop}%
\bibitem [{\citenamefont {Kun}\ \emph {et~al.}(2019)\citenamefont {Kun},
  \citenamefont {Biermann},\ and\ \citenamefont {Gergely}}]{Kun:2018zin}%
  \BibitemOpen
  \bibfield  {author} {\bibinfo {author} {\bibfnamefont {E.}~\bibnamefont
  {Kun}}, \bibinfo {author} {\bibfnamefont {P.~L.}\ \bibnamefont {Biermann}},\
  and\ \bibinfo {author} {\bibfnamefont {L.}~\bibnamefont {Gergely}},\ }\href
  {https://doi.org/10.1093/mnrasl/sly216} {\bibfield  {journal} {\bibinfo
  {journal} {Mon. Not. Roy. Astron. Soc.}\ }\textbf {\bibinfo {volume} {483}},\
  \bibinfo {pages} {L42} (\bibinfo {year} {2019})},\ \Eprint
  {https://arxiv.org/abs/1807.07942} {arXiv:1807.07942 [astro-ph.HE]}
  \BibitemShut {NoStop}%
\bibitem [{\citenamefont {{Selina}}\ \emph {et~al.}(2018)\citenamefont
  {{Selina}} \emph {et~al.}}]{2018SPIE10700E..1OS}%
  \BibitemOpen
  \bibfield  {author} {\bibinfo {author} {\bibfnamefont {R.~J.}\ \bibnamefont
  {{Selina}}} \emph {et~al.} (\bibinfo {collaboration} {ngVLA}),\ }in\ \href
  {https://doi.org/10.1117/12.2312089} {\emph {\bibinfo {booktitle}
  {Ground-based and Airborne Telescopes VII}}},\ \bibinfo {series} {Society of
  Photo-Optical Instrumentation Engineers (SPIE) Conference Series}, Vol.\
  \bibinfo {volume} {10700}\ (\bibinfo {year} {2018})\ p.\ \bibinfo {pages}
  {107001O},\ \Eprint {https://arxiv.org/abs/1806.08405} {arXiv:1806.08405
  [astro-ph.IM]} \BibitemShut {NoStop}%
\bibitem [{\citenamefont {Robertson}\ \emph {et~al.}(2017)\citenamefont
  {Robertson} \emph {et~al.}}]{Robertson:2017glv}%
  \BibitemOpen
  \bibfield  {author} {\bibinfo {author} {\bibfnamefont {B.~E.}\ \bibnamefont
  {Robertson}} \emph {et~al.} (\bibinfo {collaboration} {Members of the LSST
  Galaxies Science}),\ }\href@noop {} {\  (\bibinfo {year} {2017})},\ \Eprint
  {https://arxiv.org/abs/1708.01617} {arXiv:1708.01617 [astro-ph.GA]}
  \BibitemShut {NoStop}%
\bibitem [{\citenamefont {{Abell}}\ \emph {et~al.}(2009)\citenamefont {{Abell}}
  \emph {et~al.}}]{2009arXiv0912.0201L}%
  \BibitemOpen
  \bibfield  {author} {\bibinfo {author} {\bibfnamefont {P.~A.}\ \bibnamefont
  {{Abell}}} \emph {et~al.} (\bibinfo {collaboration} {LSST Science
  Collaboration}),\ }\href@noop {} {\bibfield  {journal} {\bibinfo  {journal}
  {arXiv e-prints}\ } (\bibinfo {year} {2009})},\ \Eprint
  {https://arxiv.org/abs/0912.0201} {arXiv:0912.0201 [astro-ph.IM]}
  \BibitemShut {NoStop}%
\bibitem [{\citenamefont {Boettcher}\ \emph {et~al.}(2013)\citenamefont
  {Boettcher}, \citenamefont {Reimer}, \citenamefont {Sweeney},\ and\
  \citenamefont {Prakash}}]{Boettcher:2013wxa}%
  \BibitemOpen
  \bibfield  {author} {\bibinfo {author} {\bibfnamefont {M.}~\bibnamefont
  {Boettcher}}, \bibinfo {author} {\bibfnamefont {A.}~\bibnamefont {Reimer}},
  \bibinfo {author} {\bibfnamefont {K.}~\bibnamefont {Sweeney}},\ and\ \bibinfo
  {author} {\bibfnamefont {A.}~\bibnamefont {Prakash}},\ }\href
  {https://doi.org/10.1088/0004-637X/768/1/54} {\bibfield  {journal} {\bibinfo
  {journal} {Astrophys. J.}\ }\textbf {\bibinfo {volume} {768}},\ \bibinfo
  {pages} {54} (\bibinfo {year} {2013})},\ \Eprint
  {https://arxiv.org/abs/1304.0605} {arXiv:1304.0605 [astro-ph.HE]}
  \BibitemShut {NoStop}%
\bibitem [{\citenamefont {{Petropoulou}}\ \emph {et~al.}(2017)\citenamefont
  {{Petropoulou}}, \citenamefont {{Vasilopoulos}},\ and\ \citenamefont
  {{Giannios}}}]{PetroVas17}%
  \BibitemOpen
  \bibfield  {author} {\bibinfo {author} {\bibfnamefont {M.}~\bibnamefont
  {{Petropoulou}}}, \bibinfo {author} {\bibfnamefont {G.}~\bibnamefont
  {{Vasilopoulos}}},\ and\ \bibinfo {author} {\bibfnamefont {D.}~\bibnamefont
  {{Giannios}}},\ }\href {https://doi.org/10.1093/mnras/stw2453} {\bibfield
  {journal} {\bibinfo  {journal} {\mnras}\ }\textbf {\bibinfo {volume} {464}},\
  \bibinfo {pages} {2213} (\bibinfo {year} {2017})},\ \Eprint
  {https://arxiv.org/abs/1608.07300} {arXiv:1608.07300 [astro-ph.HE]}
  \BibitemShut {NoStop}%
\bibitem [{\citenamefont {Evans}\ \emph {et~al.}(2015)\citenamefont {Evans}
  \emph {et~al.}}]{Evans:2015qia}%
  \BibitemOpen
  \bibfield  {author} {\bibinfo {author} {\bibfnamefont {P.~A.}\ \bibnamefont
  {Evans}} \emph {et~al.},\ }\href {https://doi.org/10.1093/mnras/stv136}
  {\bibfield  {journal} {\bibinfo  {journal} {Mon. Not. Roy. Astron. Soc.}\
  }\textbf {\bibinfo {volume} {448}},\ \bibinfo {pages} {2210} (\bibinfo {year}
  {2015})},\ \Eprint {https://arxiv.org/abs/1501.04435} {arXiv:1501.04435
  [astro-ph.HE]} \BibitemShut {NoStop}%
\bibitem [{\citenamefont {Keivani}\ \emph
  {et~al.}(2018{\natexlab{b}})\citenamefont {Keivani}, \citenamefont {Cowen},
  \citenamefont {Fox}, \citenamefont {Kennea}, \citenamefont {Tešić},
  \citenamefont {Turley}, \citenamefont {Evans}, \citenamefont {Osborne},\ and\
  \citenamefont {Marshall}}]{Keivani:2017icrc}%
  \BibitemOpen
  \bibfield  {author} {\bibinfo {author} {\bibfnamefont {A.}~\bibnamefont
  {Keivani}}, \bibinfo {author} {\bibfnamefont {D.}~\bibnamefont {Cowen}},
  \bibinfo {author} {\bibfnamefont {D.~B.}\ \bibnamefont {Fox}}, \bibinfo
  {author} {\bibfnamefont {J.}~\bibnamefont {Kennea}}, \bibinfo {author}
  {\bibfnamefont {G.}~\bibnamefont {Tešić}}, \bibinfo {author} {\bibfnamefont
  {C.~F.}\ \bibnamefont {Turley}}, \bibinfo {author} {\bibfnamefont
  {P.}~\bibnamefont {Evans}}, \bibinfo {author} {\bibfnamefont
  {J.}~\bibnamefont {Osborne}},\ and\ \bibinfo {author} {\bibfnamefont {F.~E.}\
  \bibnamefont {Marshall}},\ }\bibfield  {booktitle} {\emph {\bibinfo
  {booktitle} {{Contributions to the 35th International Cosmic Ray Conference
  (ICRC 2017)}}},\ }\href {https://doi.org/10.22323/1.301.1015} {\bibfield
  {journal} {\bibinfo  {journal} {PoS}\ }\textbf {\bibinfo {volume}
  {ICRC2017}},\ \bibinfo {pages} {1015} (\bibinfo {year}
  {2018}{\natexlab{b}})}\BibitemShut {NoStop}%
\bibitem [{\citenamefont {Aartsen}\ \emph
  {et~al.}(2017{\natexlab{b}})\citenamefont {Aartsen} \emph
  {et~al.}}]{Aartsen:2017snx}%
  \BibitemOpen
  \bibfield  {author} {\bibinfo {author} {\bibfnamefont {M.~G.}\ \bibnamefont
  {Aartsen}} \emph {et~al.},\ }\href
  {https://doi.org/10.1051/0004-6361/201730620} {\bibfield  {journal} {\bibinfo
   {journal} {Astron. Astrophys.}\ }\textbf {\bibinfo {volume} {607}},\
  \bibinfo {pages} {A115} (\bibinfo {year} {2017}{\natexlab{b}})},\ \Eprint
  {https://arxiv.org/abs/1702.06131} {arXiv:1702.06131 [astro-ph.HE]}
  \BibitemShut {NoStop}%
\bibitem [{\citenamefont {{Ubertini}}\ and\ \citenamefont
  {{Bazzano}}(2014)}]{2014NIMPA.742...47U}%
  \BibitemOpen
  \bibfield  {author} {\bibinfo {author} {\bibfnamefont {P.}~\bibnamefont
  {{Ubertini}}}\ and\ \bibinfo {author} {\bibfnamefont {A.}~\bibnamefont
  {{Bazzano}}},\ }\href {https://doi.org/10.1016/j.nima.2013.12.027} {\bibfield
   {journal} {\bibinfo  {journal} {Nuclear Instruments and Methods in Physics
  Research A}\ }\textbf {\bibinfo {volume} {742}},\ \bibinfo {pages} {47}
  (\bibinfo {year} {2014})}\BibitemShut {NoStop}%
\bibitem [{\citenamefont {Kawamuro}\ \emph {et~al.}(2018)\citenamefont
  {Kawamuro} \emph {et~al.}}]{Kawamuro:2018eky}%
  \BibitemOpen
  \bibfield  {author} {\bibinfo {author} {\bibfnamefont {T.}~\bibnamefont
  {Kawamuro}} \emph {et~al.},\ }\href
  {https://doi.org/10.3847/1538-4365/aad1ef} {\bibfield  {journal} {\bibinfo
  {journal} {Astrophys. J. Suppl.}\ }\textbf {\bibinfo {volume} {238}},\
  \bibinfo {pages} {32} (\bibinfo {year} {2018})},\ \Eprint
  {https://arxiv.org/abs/1807.00874} {arXiv:1807.00874 [astro-ph.HE]}
  \BibitemShut {NoStop}%
\bibitem [{\citenamefont {{Camp}}\ and\ \citenamefont {{Transient Astrophysics
  Probe Team}}(2018)}]{2018AAS...23112105C}%
  \BibitemOpen
  \bibfield  {author} {\bibinfo {author} {\bibfnamefont {J.}~\bibnamefont
  {{Camp}}}\ and\ \bibinfo {author} {\bibnamefont {{Transient Astrophysics
  Probe Team}}},\ }in\ \href@noop {} {\emph {\bibinfo {booktitle} {American
  Astronomical Society Meeting Abstracts \#231}}},\ \bibinfo {series} {American
  Astronomical Society Meeting Abstracts}, Vol.\ \bibinfo {volume} {231}\
  (\bibinfo {year} {2018})\ p.\ \bibinfo {pages} {121.05}\BibitemShut {NoStop}%
\bibitem [{\citenamefont {Ray}\ \emph {et~al.}(2018)\citenamefont {Ray} \emph
  {et~al.}}]{Ray:2018dlb}%
  \BibitemOpen
  \bibfield  {author} {\bibinfo {author} {\bibfnamefont {P.~S.}\ \bibnamefont
  {Ray}} \emph {et~al.},\ }\bibfield  {booktitle} {\emph {\bibinfo {booktitle}
  {{Proceedings, SPIE Astronomical Telescopes + Instrumentation 2018: Modeling,
  Systems Engineering, and Project Management for Astronomy VIII: Austin, USA,
  June 10-15, 2018}}},\ }\href {https://doi.org/10.1117/12.2312257} {\bibfield
  {journal} {\bibinfo  {journal} {Proc. SPIE Int. Soc. Opt. Eng.}\ }\textbf
  {\bibinfo {volume} {10699}},\ \bibinfo {pages} {1069919} (\bibinfo {year}
  {2018})},\ \Eprint {https://arxiv.org/abs/1807.01179} {arXiv:1807.01179
  [astro-ph.IM]} \BibitemShut {NoStop}%
\bibitem [{\citenamefont {{Cordier}}\ \emph {et~al.}(2015)\citenamefont
  {{Cordier}} \emph {et~al.}}]{2015arXiv151203323C}%
  \BibitemOpen
  \bibfield  {author} {\bibinfo {author} {\bibfnamefont {B.}~\bibnamefont
  {{Cordier}}} \emph {et~al.},\ }\href@noop {} {\bibfield  {journal} {\bibinfo
  {journal} {arXiv e-prints}\ } (\bibinfo {year} {2015})},\ \Eprint
  {https://arxiv.org/abs/1512.03323} {arXiv:1512.03323 [astro-ph.IM]}
  \BibitemShut {NoStop}%
\bibitem [{\citenamefont {{Fox}}\ \emph {et~al.}(2017)\citenamefont {{Fox}},
  \citenamefont {{DeLaunay}}, \citenamefont {{Keivani}}, \citenamefont
  {{Evans}}, \citenamefont {{Turley}}, \citenamefont {{Kennea}}, \citenamefont
  {{Cowen}}, \citenamefont {{Osborne}}, \citenamefont {{Santander}},\ and\
  \citenamefont {{Marshall}}}]{2017ATel10845....1F}%
  \BibitemOpen
  \bibfield  {author} {\bibinfo {author} {\bibfnamefont {D.~B.}\ \bibnamefont
  {{Fox}}}, \bibinfo {author} {\bibfnamefont {J.~J.}\ \bibnamefont
  {{DeLaunay}}}, \bibinfo {author} {\bibfnamefont {A.}~\bibnamefont
  {{Keivani}}}, \bibinfo {author} {\bibfnamefont {P.~A.}\ \bibnamefont
  {{Evans}}}, \bibinfo {author} {\bibfnamefont {C.~F.}\ \bibnamefont
  {{Turley}}}, \bibinfo {author} {\bibfnamefont {J.~A.}\ \bibnamefont
  {{Kennea}}}, \bibinfo {author} {\bibfnamefont {D.~F.}\ \bibnamefont
  {{Cowen}}}, \bibinfo {author} {\bibfnamefont {J.~P.}\ \bibnamefont
  {{Osborne}}}, \bibinfo {author} {\bibfnamefont {M.}~\bibnamefont
  {{Santander}}},\ and\ \bibinfo {author} {\bibfnamefont {F.~E.}\ \bibnamefont
  {{Marshall}}},\ }\href@noop {} {\bibfield  {journal} {\bibinfo  {journal}
  {The Astronomer's Telegram}\ }\textbf {\bibinfo {volume} {10845}} (\bibinfo
  {year} {2017})}\BibitemShut {NoStop}%
\bibitem [{\citenamefont {Moiseev}\ \emph {et~al.}(2018)\citenamefont {Moiseev}
  \emph {et~al.}}]{Moiseev:2017mxg}%
  \BibitemOpen
  \bibfield  {author} {\bibinfo {author} {\bibfnamefont {A.}~\bibnamefont
  {Moiseev}} \emph {et~al.} (\bibinfo {collaboration} {AMEGO}),\ }\bibfield
  {booktitle} {\emph {\bibinfo {booktitle} {{Contributions to the 35th
  International Cosmic Ray Conference (ICRC 2017)}}},\ }\href
  {https://doi.org/10.22323/1.301.0798} {\bibfield  {journal} {\bibinfo
  {journal} {PoS}\ }\textbf {\bibinfo {volume} {ICRC2017}},\ \bibinfo {pages}
  {798} (\bibinfo {year} {2018})}\BibitemShut {NoStop}%
\bibitem [{\citenamefont {Zhang}\ and\ \citenamefont
  {Böttcher}(2013)}]{Zhang:2013bna}%
  \BibitemOpen
  \bibfield  {author} {\bibinfo {author} {\bibfnamefont {H.}~\bibnamefont
  {Zhang}}\ and\ \bibinfo {author} {\bibfnamefont {M.}~\bibnamefont
  {Böttcher}},\ }\href {https://doi.org/10.1088/0004-637X/774/1/18} {\bibfield
   {journal} {\bibinfo  {journal} {Astrophys. J.}\ }\textbf {\bibinfo {volume}
  {774}},\ \bibinfo {pages} {18} (\bibinfo {year} {2013})},\ \Eprint
  {https://arxiv.org/abs/1307.4187} {arXiv:1307.4187 [astro-ph.HE]}
  \BibitemShut {NoStop}%
\bibitem [{\citenamefont {Paliya}\ \emph {et~al.}(2018)\citenamefont {Paliya},
  \citenamefont {Zhang}, \citenamefont {Böttcher}, \citenamefont {Ajello},
  \citenamefont {Domínguez}, \citenamefont {Joshi}, \citenamefont {Hartmann},\
  and\ \citenamefont {Stalin}}]{Paliya:2018wgu}%
  \BibitemOpen
  \bibfield  {author} {\bibinfo {author} {\bibfnamefont {V.~S.}\ \bibnamefont
  {Paliya}}, \bibinfo {author} {\bibfnamefont {H.}~\bibnamefont {Zhang}},
  \bibinfo {author} {\bibfnamefont {M.}~\bibnamefont {Böttcher}}, \bibinfo
  {author} {\bibfnamefont {M.}~\bibnamefont {Ajello}}, \bibinfo {author}
  {\bibfnamefont {A.}~\bibnamefont {Domínguez}}, \bibinfo {author}
  {\bibfnamefont {M.}~\bibnamefont {Joshi}}, \bibinfo {author} {\bibfnamefont
  {D.}~\bibnamefont {Hartmann}},\ and\ \bibinfo {author} {\bibfnamefont
  {C.~S.}\ \bibnamefont {Stalin}},\ }\href
  {https://doi.org/10.3847/1538-4357/aad1f0} {\bibfield  {journal} {\bibinfo
  {journal} {Astrophys. J.}\ }\textbf {\bibinfo {volume} {863}},\ \bibinfo
  {pages} {98} (\bibinfo {year} {2018})},\ \Eprint
  {https://arxiv.org/abs/1807.02085} {arXiv:1807.02085 [astro-ph.HE]}
  \BibitemShut {NoStop}%
\bibitem [{\citenamefont {{Zhang}}\ \emph {et~al.}(2019)\citenamefont
  {{Zhang}}, \citenamefont {{Fang}}, \citenamefont {{Li}}, \citenamefont
  {{Giannios}}, \citenamefont {{B{\"o}ttcher}},\ and\ \citenamefont
  {{Buson}}}]{Zhang:2018xrr}%
  \BibitemOpen
  \bibfield  {author} {\bibinfo {author} {\bibfnamefont {H.}~\bibnamefont
  {{Zhang}}}, \bibinfo {author} {\bibfnamefont {K.}~\bibnamefont {{Fang}}},
  \bibinfo {author} {\bibfnamefont {H.}~\bibnamefont {{Li}}}, \bibinfo {author}
  {\bibfnamefont {D.}~\bibnamefont {{Giannios}}}, \bibinfo {author}
  {\bibfnamefont {M.}~\bibnamefont {{B{\"o}ttcher}}},\ and\ \bibinfo {author}
  {\bibfnamefont {S.}~\bibnamefont {{Buson}}},\ }\href@noop {} {\bibfield
  {journal} {\bibinfo  {journal} {arXiv e-prints}\ ,\ \bibinfo {eid}
  {arXiv:1903.01956}} (\bibinfo {year} {2019})},\ \Eprint
  {https://arxiv.org/abs/1903.01956} {arXiv:1903.01956 [astro-ph.HE]}
  \BibitemShut {NoStop}%
\bibitem [{\citenamefont {Weisskopf}\ \emph {et~al.}(2016)\citenamefont
  {Weisskopf} \emph {et~al.}}]{WEISSKOPF20161179}%
  \BibitemOpen
  \bibfield  {author} {\bibinfo {author} {\bibfnamefont {M.~C.}\ \bibnamefont
  {Weisskopf}} \emph {et~al.} (\bibinfo {collaboration} {IXPE}),\ }\href
  {https://doi.org/https://doi.org/10.1016/j.rinp.2016.10.021} {\bibfield
  {journal} {\bibinfo  {journal} {Results in Physics}\ }\textbf {\bibinfo
  {volume} {6}},\ \bibinfo {pages} {1179 } (\bibinfo {year}
  {2016})}\BibitemShut {NoStop}%
\bibitem [{\citenamefont {{Atwood}}\ \emph {et~al.}(2009)\citenamefont
  {{Atwood}} \emph {et~al.}}]{2009ApJ...697.1071A}%
  \BibitemOpen
  \bibfield  {author} {\bibinfo {author} {\bibfnamefont {W.~B.}\ \bibnamefont
  {{Atwood}}} \emph {et~al.} (\bibinfo {collaboration} {Fermi-LAT}),\ }\href
  {https://doi.org/10.1088/0004-637X/697/2/1071} {\bibfield  {journal}
  {\bibinfo  {journal} {\apj}\ }\textbf {\bibinfo {volume} {697}},\ \bibinfo
  {pages} {1071} (\bibinfo {year} {2009})},\ \Eprint
  {https://arxiv.org/abs/0902.1089} {arXiv:0902.1089 [astro-ph.IM]}
  \BibitemShut {NoStop}%
\bibitem [{\citenamefont {Santander}(2018)}]{Santander:2017zkl}%
  \BibitemOpen
  \bibfield  {author} {\bibinfo {author} {\bibfnamefont {M.}~\bibnamefont
  {Santander}} (\bibinfo {collaboration} {VERITAS, FACT, IceCube, MAGIC,
  H.E.S.S.}),\ }\bibfield  {booktitle} {\emph {\bibinfo {booktitle}
  {{Contributions to the 35th International Cosmic Ray Conference (ICRC
  2017)}}},\ }\href {https://doi.org/10.22323/1.301.0618} {\bibfield  {journal}
  {\bibinfo  {journal} {PoS}\ }\textbf {\bibinfo {volume} {ICRC2017}},\
  \bibinfo {pages} {618} (\bibinfo {year} {2018})},\ \bibinfo {note}
  {[35,618(2017)]},\ \Eprint {https://arxiv.org/abs/1708.08945}
  {arXiv:1708.08945 [astro-ph.HE]} \BibitemShut {NoStop}%
\bibitem [{\citenamefont {Acharya}\ \emph {et~al.}(2018)\citenamefont {Acharya}
  \emph {et~al.}}]{Acharya:2017ttl}%
  \BibitemOpen
  \bibfield  {author} {\bibinfo {author} {\bibfnamefont {B.~S.}\ \bibnamefont
  {Acharya}} \emph {et~al.} (\bibinfo {collaboration} {CTA Consortium}),\
  }\href {https://doi.org/10.1142/10986} {\emph {\bibinfo {title} {{Science
  with the Cherenkov Telescope Array}}}}\ (\bibinfo  {publisher} {WSP},\
  \bibinfo {year} {2018})\ \Eprint {https://arxiv.org/abs/1709.07997}
  {arXiv:1709.07997 [astro-ph.IM]} \BibitemShut {NoStop}%
\bibitem [{\citenamefont {Abeysekara}\ \emph
  {et~al.}(2017{\natexlab{b}})\citenamefont {Abeysekara} \emph
  {et~al.}}]{Abeysekara:2017hyn}%
  \BibitemOpen
  \bibfield  {author} {\bibinfo {author} {\bibfnamefont {A.~U.}\ \bibnamefont
  {Abeysekara}} \emph {et~al.} (\bibinfo {collaboration} {HAWC}),\ }\href
  {https://doi.org/10.3847/1538-4357/aa7556} {\bibfield  {journal} {\bibinfo
  {journal} {Astrophys. J.}\ }\textbf {\bibinfo {volume} {843}},\ \bibinfo
  {pages} {40} (\bibinfo {year} {2017}{\natexlab{b}})},\ \Eprint
  {https://arxiv.org/abs/1702.02992} {arXiv:1702.02992 [astro-ph.HE]}
  \BibitemShut {NoStop}%
\bibitem [{\citenamefont {Chen}(2018)}]{Chen:2018glo}%
  \BibitemOpen
  \bibfield  {author} {\bibinfo {author} {\bibfnamefont {M.}~\bibnamefont
  {Chen}} (\bibinfo {collaboration} {LHAASO}),\ }\bibfield  {booktitle} {\emph
  {\bibinfo {booktitle} {{Contributions to the 35th International Cosmic Ray
  Conference (ICRC 2017)}}},\ }\href {https://doi.org/10.22323/1.301.0832}
  {\bibfield  {journal} {\bibinfo  {journal} {PoS}\ }\textbf {\bibinfo {volume}
  {ICRC2017}},\ \bibinfo {pages} {832} (\bibinfo {year} {2018})}\BibitemShut
  {NoStop}%
\bibitem [{\citenamefont {Assis}\ \emph {et~al.}(2017)\citenamefont {Assis}
  \emph {et~al.}}]{Assis:2017zzn}%
  \BibitemOpen
  \bibfield  {author} {\bibinfo {author} {\bibfnamefont {P.}~\bibnamefont
  {Assis}} \emph {et~al.} (\bibinfo {collaboration} {LATTES}),\ }\bibfield
  {booktitle} {\emph {\bibinfo {booktitle} {{Proceedings, 6th Roma
  International Workshop on Astroparticle Physics (RICAP16): Rome, Italy, June
  21-24, 2016}}},\ }\href {https://doi.org/10.1051/epjconf/201713603013,
  10.1051/epjconf/e2016-62145-8} {\bibfield  {journal} {\bibinfo  {journal}
  {EPJ Web Conf.}\ }\textbf {\bibinfo {volume} {136}},\ \bibinfo {pages}
  {03013} (\bibinfo {year} {2017})},\ \Eprint
  {https://arxiv.org/abs/1703.09254} {arXiv:1703.09254 [astro-ph.IM]}
  \BibitemShut {NoStop}%
\bibitem [{\citenamefont {Thoudam}\ \emph {et~al.}(2018)\citenamefont
  {Thoudam}, \citenamefont {Becherini},\ and\ \citenamefont
  {Punch}}]{Thoudam:2017hgj}%
  \BibitemOpen
  \bibfield  {author} {\bibinfo {author} {\bibfnamefont {S.}~\bibnamefont
  {Thoudam}}, \bibinfo {author} {\bibfnamefont {Y.}~\bibnamefont {Becherini}},\
  and\ \bibinfo {author} {\bibfnamefont {M.}~\bibnamefont {Punch}} (\bibinfo
  {collaboration} {ALTO}),\ }\bibfield  {booktitle} {\emph {\bibinfo
  {booktitle} {{Contributions to the 35th International Cosmic Ray Conference
  (ICRC 2017)}}},\ }\href {https://doi.org/10.22323/1.301.0780} {\bibfield
  {journal} {\bibinfo  {journal} {PoS}\ }\textbf {\bibinfo {volume}
  {ICRC2017}},\ \bibinfo {pages} {780} (\bibinfo {year} {2018})},\ \bibinfo
  {note} {[35,780(2017)]},\ \Eprint {https://arxiv.org/abs/1708.01059}
  {arXiv:1708.01059 [astro-ph.IM]} \BibitemShut {NoStop}%
\bibitem [{\citenamefont {Asaba}\ \emph {et~al.}(2018)\citenamefont {Asaba}
  \emph {et~al.}}]{Ohnishi:2017qsz}%
  \BibitemOpen
  \bibfield  {author} {\bibinfo {author} {\bibfnamefont {T.}~\bibnamefont
  {Asaba}} \emph {et~al.} (\bibinfo {collaboration} {ALPACA}),\ }\bibfield
  {booktitle} {\emph {\bibinfo {booktitle} {{Contributions to the 35th
  International Cosmic Ray Conference (ICRC 2017)}}},\ }\href
  {https://doi.org/10.22323/1.301.0827} {\bibfield  {journal} {\bibinfo
  {journal} {PoS}\ }\textbf {\bibinfo {volume} {ICRC2017}},\ \bibinfo {pages}
  {827} (\bibinfo {year} {2018})}\BibitemShut {NoStop}%
\end{thebibliography}%
\end{document}